\documentclass[aps,prd,a4paper,twocolumn,amsmath,showpacs,superscriptaddress,nofootinbib,preprintnumbers]{revtex4-1}

\usepackage{verbatim}
\usepackage[T1]{fontenc}
\usepackage[utf8]{inputenc}
\usepackage[american]{babel}
\usepackage{epsfig}
\usepackage{graphicx}
\usepackage{booktabs}
\usepackage{multirow}
\usepackage{dcolumn}
\usepackage{amsmath}
\usepackage{mathtools}
\usepackage{amsfonts}
\usepackage{amssymb}
\usepackage{epstopdf}
\usepackage{bm}
\usepackage{siunitx}
\usepackage{braket}
\usepackage{enumitem}
\usepackage{soul}
\usepackage[table]{xcolor}
\usepackage{color}
\usepackage{transparent}
\usepackage{pifont}
\definecolor{navyblue}{rgb}{0.0, 0.0, 0.5}
\definecolor{royalblue}{rgb}{0.25, 0.41, 0.88}
\definecolor{cadmiumgreen}{rgb}{0.0, 0.42, 0.24}
\definecolor{blue-violet}{rgb}{0.54, 0.17, 0.89}
\definecolor{darkviolet}{rgb}{0.58, 0.0, 0.83}
\definecolor{orange(colorwheel)}{rgb}{1.0, 0.5, 0.0}

\usepackage{hyperref}
\hypersetup{
    colorlinks=true, 
    linkcolor=royalblue, 
    citecolor=magenta}

\newcommand\ee{\end{equation}}
\newcommand\be{\begin{equation}}
\newcommand\eea{\end{eqnarray}}
\newcommand\bea{\begin{eqnarray}}

\renewcommand\({\left(}
\renewcommand\){\right)}
\renewcommand\[{\left[}
\renewcommand\]{\right]}

\usepackage{booktabs}
\usepackage{multirow}
\usepackage{dcolumn}
\usepackage{colortbl}
\newcommand\vertsp{\rule[-2mm]{1mm}{0mm} &}
\newcommand\horsp{\rule[-1.5mm]{0mm}{4.125mm}}
\newcommand\morehorsp{\rule[-2.25mm]{0mm}{6mm}}
\newcommand{\siround}[2]{\num[round-mode=places,group-digits=false,round-precision=#2]{#1}}

\definecolor{magenta(process)}{rgb}{1.0, 0.0, 0.56}

\definecolor{darkspringgreen}{rgb}{0.09, 0.45, 0.27}

\definecolor{royalblue(web)}{rgb}{0.25, 0.41, 0.88}

\begin{document}

\title{The cosmological impact of future constraints on $H_0$ from gravitational-wave standard sirens}

\author{Eleonora Di Valentino}
\email{eleonora.divalentino@manchester.ac.uk}
\affiliation{Jodrell Bank Center for Astrophysics, School of Physics and Astronomy, University of Manchester, Oxford Road, Manchester, M13 9PL, UK}

\author{Daniel E. Holz}
\email{holz@uchicago.edu}
\affiliation{Enrico Fermi Institute, Department of Physics, Department of Astronomy and Astrophysics,\\and Kavli Institute for Cosmological Physics, University of Chicago, Chicago, IL 60637, USA}
\affiliation{Kavli Institute for Particle Astrophysics \& Cosmology and Physics Department,\\Stanford University, Stanford, CA 94305}

\author{Alessandro Melchiorri}
\email{alessandro.melchiorri@roma1.infn.it}
\affiliation{Physics Department and INFN, Universit\`a di Roma ``La Sapienza'', Ple Aldo Moro 2, 00185, Rome, Italy} 

\author{Fabrizio Renzi}
\email{fabrizio.renzi@roma1.infn.it}
\affiliation{Physics Department and INFN, Universit\`a di Roma ``La Sapienza'', Ple Aldo Moro 2, 00185, Rome, Italy}

\date{\today}

\preprint{}
\begin{abstract}
Gravitational-wave standard sirens present a novel approach for the determination of the Hubble constant. After the recent spectacular confirmation of the method thanks to GW170817 and its optical counterpart, additional standard siren measurements from future gravitational-wave sources are expected to constrain the Hubble constant to high accuracy. At the same time, improved constraints are expected from observations of cosmic microwave background (CMB) polarization and from baryon acoustic oscillations (BAO) surveys. We explore the role of future standard siren constraints on $H_0$ in light of expected CMB+BAO data. 
Considering a $10$-parameters cosmological model, in which curvature, the dark energy equation of state, and the Hubble constant are unbounded by CMB observations, we find that a combination of future CMB+BAO data will constrain the Hubble parameter to $\sim 1.5 \%$. Further extending the parameter space to a time-varying dark energy equation of state, we find that future CMB+BAO constraints on $H_0$ are relaxed to $\sim 3.0 \%$. These accuracies are within reach of future standard siren measurements from the Hanford-Livingston-Virgo and  the Hanford-Livingston-Virgo-Japan-India networks of interferometers, showing the cosmological relevance of these sources. If future gravitational-wave standard siren measurements reach $1\%$ on $H_0$, as expected, they would significantly improve future CMB+BAO constraints on curvature and on the dark energy equation of state by up to a factor $\sim 3$. We also show that the inclusion of $H_0$ constraints from gravitational-wave standard sirens could result in a reduction of the dark energy figure-of-merit (i.e., the cosmological parameter volume) by up to a factor of $\sim 400$. 

\end{abstract}

\maketitle

\section{Introduction}

Gravitational-wave standard sirens (GWSS) have been proposed as a powerful method for the determination of the Hubble constant (see e.g. \cite{schutz,Holz2005,gwss1, nissanke,holz,ref1,ref2,ref3,ref4}).  The feasibility of the method has been experimentally confirmed by the recent spectacular observations of the event GW170817 \cite{GW170817} and the detection of an associated optical counterpart \cite{2017Sci...358.1556C,2017ApJ...848L..16S,2017ApJ...848L..12A}, yielding a constraint of $H_0=70_{-8}^{+12}$ km/s/Mpc (maximum a posteriori value with minimal 68.3\% credible interval) \cite{natureH0}.
While this constraint is much weaker than those currently obtained from measurements of luminosity distances of standard sirens or observations of the cosmic microwave background (CMB) anisotropies, it is expected to significantly improve in the coming years with the discovery of additional standard siren events. 
Moreover, this kind of measurement is clearly of particular interest given the current discrepancy on the value of $H_0$ between standard candle luminosity distances of Cepheids and Type Ia supernovae, that report a value of $H_0=73.24\pm 1.74$ km/s/Mpc at $68 \%$ C.L. \cite{R16} (R16, hereafter), 
($H_0=73.52\pm 1.62$ km/s/Mpc at $68 \%$ C.L. in the new analysis of \cite{R18}), and CMB measurements from the latest Planck satellite 2018 release  that gives $H_0=67.27\pm0.60$ km/s/Mpc at $68 \%$ C.L. (\cite{Aghanim:2018eyx}, see also \cite{plancknewtau, freedman,planck2015}). Current observations of baryon acoustic oscillations (BAO) are in agreement with the Planck cosmology and a combined Planck+BAO analysis gives $H_0=67.67\pm0.45$ km/s/Mpc at $68 \%$ C.L. \cite{Aghanim:2018eyx}, also in strong discrepancy with the standard candle results of \cite{R16,R18}.

While unidentified systematics could be clearly present, this tension may indicate the need of new physics beyond the standard $\Lambda$CDM model. Indeed, since the Planck constraint is derived under the assumption of $\Lambda$CDM, simple extensions to this model could relax the CMB constraint. For example, new physics in the dark energy or  neutrino sectors can significantly undermine the Planck constraints on the Hubble constant, solving the current tension on $H_0$ (see e.g. \cite{R16,papero,paper1,paper2,
Zhao:2017urm,Yang:2017amu,Prilepina:2016rlq,Santos:2016sog,Kumar:2016zpg,Karwal:2016vyq,benetti,Ko:2016uft,Archidiacono:2016kkh,Qing-Guo:2016ykt,Zhang:2017idq,zhao,Sola:2017jbl,brust,mena,boehm,wekiang,paper0}). Moreover, a time varying dark energy equation of state could also alleviate the tension between the Planck+BAO and the R16 constraint (see e.g. \cite{paper2,Zhao:2017cud,DiValentino:2017gzb}). 
Clearly an independent and accurate future determination of $H_0$ from GWSS will play a key role in confirming or rejecting the possibility of new physics beyond $\Lambda$CDM. 

It is to emphasized that an accurate measurement of the Hubble constant, even though it is a low-redshift quantity, can have important consequences for other higher-redshift cosmological parameters such as the dark energy equation of state~\cite{2005ASPC..339..215H}.
The possibility of constraining cosmology with GWSS has been already considered in several previous works (see e.g. \cite{gwss1,gwss2,gwss3,gwss4,gwss5,gwss5,gwss6,gwss7,gwss8,gwss9,gwss10,gwss11}). Some of these studies analyzed ``far future'' experiments such as the LISA satellite mission \cite{lisa} expected to be launched in $2034$ or third generation interferometers such as the Einstein Telescope \cite{ET} or the Cosmic Explorer \cite{CE}. However, recently, in \cite{holz} it has been estimated that, depending on the discovery rate of binary neutron stars, a sub-percent determination of the Hubble constant from GWSS could be achieved by the Hanford-Livingston-Virgo (HLV) network as early as during the second year of operation at design sensitivity ($\sim2023$~\cite{2018LRR....21....3A}). Given the rate uncertainties, a sub-percent measurement may have to wait for two years of the Hanford-Livingston-Virgo-Japan-India (HLVJI) network, which is expected to commence operations $\sim 2024+$. 
On the other hand, a significant improvement in the observational data is expected from the next CMB and BAO experiments. Future satellite missions such as LiteBIRD \cite{litebird} and ground based experiments such as CMB-S4 \cite{stage4} will improve the Planck results thanks to cosmic variance limited measurements of CMB polarization. The LiteBIRD satellite is a JAXA strategic large mission candidate in Phase-A1 (concept development) and is currently scheduled for launch around  2026--2027. A complementary ground-based CMB experiment with the sensitivity of CMB-S4 is at the moment planned after $2023$. Similarly, galaxy spectroscopic surveys such as DESI (\cite{Levi:2013gra}, expected to be completed by $2023$) will observe BAO with unprecedented precision.

The level of accuracy on the Hubble constant expected from future CMB+BAO observations can reach the $0.15 \%$ level (see e.g. \cite{core1}). This could naively appear as an order of magnitude more accurate than future projections for standard sirens constraints. However the CMB+BAO constraint is obtained under the assumption of $\Lambda$CDM and, as we show below, can easily be more than one order of magnitude weaker in extended cosmological scenarios. These extended scenarios are of particular interest as they may offer a solution to the existing tension between different measurements of $H_0$.

We emphasize that standard sirens constitute a direct measurement of the luminosity distance, obviating the need for a distance ladder. The absolute calibration of the source is provided by the theory of general relativity. The possible systematics associated with standard siren measurements are expected to reside primarily with the instrument, and in particular, with the calibration of the photodetectors which lead directly to the measurement of the amplitude of the gravitational waves \cite{abbott1,abbott2}. This calibration is expected to be achieved to better than 1\% in the near future~\cite{2016RScI...87k4503K}. Gravitational wave standard siren measurements thus have the potential to provide a particularly clean and robust probe to the sub-percent level. This is to be compared with the case of Type Ia supernovae standard candle measurements, which involve astronomical calibrators such as Cepheids, and multiple rungs of the distance ladder. It remains unclear whether the supernova systematics can be reduced to the $\sim1$\% level~(see, e.g., \cite{2016ApJ...826...56R,2017ApJ...848...56U,2018RPPh...81a6901H}). However, if supernovae achieve this level of accuracy on the measurement of $H_0$, then our results apply directly to them as well. Of course, supernovae also offer the opportunity to probe to much higher redshifts than GWSS, and therefore offer additional cosmological constraints.

It is therefore timely to investigate what kind of additional constraints a direct determination of $H_0$ with $\sim 1\%$ accuracy from GWSS can bring, with the expected completion of new CMB and BAO surveys within the coming decade.
In this paper we address this question by forecasting the cosmological constraints from future CMB and BAO surveys in extended cosmological scenarios and by discussing the implications of an additional independent and direct $H_0$ measurement at the level of $ 1\%$ from upcoming GWSS sources.

Several previous works have presented forecasts on $H_0$ from a variety of potential future cosmological datasets. Most notably, Weinberg et al. 2013 \cite{weinberg} performed a thorough analysis of future CMB, BAO, weak lensing, and supernovae data, and presented future constraint on dynamical dark energy and explicitly discussed the impact of a future $H_0$ prior. In this paper we complement and, in some cases, extend these studies; in detail:

\begin{itemize}

\item While most of the previous forecasts have adopted a Fisher Matrix approach,  we base our analysis on a Monte Carlo Markov Chain method. This is needed when the posterior distribution of the parameters is strongly non-Gaussian. As we discuss later in this paper this is the case for several key parameters when CMB and CMB+GWSS datasets are considered.

\item We use an extended $10$-parameter space including not only dynamical dark energy but also possible variations in neutrino masses and in the neutrino effective number. We also discuss the impact of the improvement of the $H_0$ prior on each of the parameters, as well as on the global Figure of Merit (FoM, hereafter).

\item We consider $4$ future, post-Planck, CMB experiments (one satellite and three ground-based) discussing the relative advantages and disadvantages of these missions. This is the first time that a similar comparison is presented in these extended parameter space
(extended parameter spaces considering the CORE-M5 proposal have been already studied in \cite{core1}.

\item In addition to the CMB data we conservatively adopt a single additional cosmological probe, namely a BAO dataset from the DESI experiment; we do not incorporate future supernovae or weak lensing measurements. The main goal of this paper is to demonstrate the kind of improvement a GWSS measurement could bring to a very conservative framework, therefore considering at the same time the largest number of parameters and the smallest number of datasets in order to minimize the presence of theoretical biases and experimental systematics. We choose CMB and BAO data since they should be, in principle, less affected by theoretical and experimental systematics (see e.g. Table I and the discussion in \cite{2018RPPh...81a6901H}) letting us to produce more accurate forecasts \footnote{For example, while future weak lensing measurements from large surveys as EUCLID (see e.g. \cite{euclid}) are extremely promising, systematic errors could limit an accurate determination of the galaxy shapes (see e.g. \cite{mandelbaum}) and redshifts (\cite{huterer2}). The accurate description of non-linearities and non-Gaussianities in extended scenarios could also become a relevant issue since at the moment most of the current predictions are computed from N-Body simulations that assume $\Lambda$CDM. An accurate modeling of all these systematics for a given experimental configuration is currently under study from the weak lensing community and it goes beyond the scope of this work. We therefore do not consider cosmic shear data in our study.}. We complement these measurements with the standard siren measurements, which enjoy a similar level of theoretical and experimental purity.
\end{itemize}

Our paper is structured as follows: in the next section we discuss our methods, in section III we present our results, and in section IV we present our conclusions.

\section{Method}

In this section we describe our forecasting method. We start with a description of the assumed theoretical framework, and then discuss the generation of forecasts for CMB, BAO, and standard sirens constraints.

\subsection{Extended models}

As discussed in the introduction, in this paper we consider parameter extensions to the standard $\Lambda$CDM model. These models, as we discuss below, are physically plausible, compatible with current observations, and able to solve in some cases the current observed tensions between cosmological datasets.
The standard flat $\Lambda$CDM model is based on just $6$ parameters: the baryon $\omega_b$ and cold dark matter $\omega_c$ physical energy densities, the amplitude $A_S$ and the spectral index $n_S$ of scalar primordial perturbations, the angular size of the sound horizon at decoupling $\theta_s$ and the optical depth at reionization $\tau$. Following \cite{paper0, DiValentino:2017clw} we consider variations with the addition of $4$ additional parameters:

\begin{itemize}

\item Curvature, $\Omega_k$. Most of recent analyses assume a flat universe with $\Omega_k=0$ since this is considered as one of the main predictions of inflation. However, inflationary models with non-zero curvature can be conceived (see e.g. \cite{linde}). Moreover the recent results from Planck prefer a closed model $\Omega_k>0$ at more than two standard deviations \cite{planck2015}. Including further data from BAO strongly constrains curvature with $\Omega_k=0.0002 \pm 0.0021$ at $68 \%$ C.L. and perfectly compatible with a flat universe \cite{planck2015}. However this result is obtained in the framework of $\Lambda$CDM+$\Omega_k$, i.e. in one single parameter extension while here we want to analyze a larger parameter space, varying ten parameters at the same time. In this scenario the current Planck+BAO constraints on $\Omega_k$ are weaker.

\item Neutrino mass, $\Sigma m_{\nu}$.  Neutrino oscillation experiments have demonstrated that neutrinos undergo flavor oscillations and must therefore have small but non-zero masses. However the neutrino absolute mass scale and the mass hierarchy are not yet determined (see e.g. \cite{capozzi} for a recent review). Usually, as in \cite{planck2015}, the total neutrino mass scale is fixed to $\Sigma m_{\nu}=0.06$eV, corresponding to the minimal value expected in the normal hierarchy scenario. There is clearly no fundamental reason to limit current analyses to this value and the neutrino mass should be let free to vary.

\item Neutrino effective number, $N_{\rm eff}$. Any particle that decouples from the primordial thermal plasma before the QCD transition could change the number of relativistic particles at recombination increasing $N_{\rm eff}$ from its standard value of $3.046$ (see e.g. \cite{baumaneff}). An increased value of $N_{\rm eff}$ can help in solving the Hubble constant tension (see e.g. \cite{R16}). 
Reheating at energy scales close to the epoch of neutrino decoupling could on the contrary lower the value of $N_{\rm eff}$ \cite{deSalas:2015glj}. 

\item Dark energy equation of state $w$. While current data are in agreement with a cosmological constant, the possibility of having a dark energy equation of state different from $-1$ is certainly open (see e.g. \cite{paper1}). Moreover, a time evolution for $w$ helps in solving the coincidence problem of why dark energy and dark matter have similar densities today. In this paper we consider two parametrizations, either $w$ constant with time or the Chevalier-Polarski-Linder parametrization (hereafter CPL) \cite{chevalier,linder}:
\begin{equation}
w(a)=w_0+(1-a)w_a
\end{equation} 
where $a$ is the scale factor, $w_0$ is the equation of state today ($a=1$) and $w_a$ parametrizes its time evolution. This should be considered as a minimal extension since dark energy time dependences could be more complicated as, for example, in the case of rapid transitions. We consider dark energy perturbations following the approach of \cite{waynehu}.
 	
\end{itemize}

In this paper we consider the following $10$ parameters extensions: $\Lambda$CDM+$\Omega_k$+$N_{\rm eff}$+$\Sigma m_{\nu}$+$w$ and $\Lambda$CDM+$\Omega_k$+$\Sigma m_{\nu}$+$w_0$+$w_a$. 
While we study extended models, for our simulated data we assume as a fiducial (true) model the standard $\Lambda$CDM model with parameters in agreement with the recent Planck constraints \cite{planck2015}: $\omega_b=0.02225$,  $\omega_{c}= 0.1198$, $\tau=0.055$, $100\theta_{MC}=1.04077$, $\Sigma m_{\nu}=0.06$ eV and $n_s=0.9645$. The corresponding derived value of $H_0$ in this model is
$H_0=67.3$ km/s/Mpc. The theoretical models and the simulated data are computed with the latest version of the Boltzmann integrator CAMB \cite{camb}. Given a simulated dataset and a likelihood that compares data with theory, we extract the constraints on cosmological parameters using the Monte Carlo Markow Chain (MCMC) code {\sc CosmoMC}\footnote{\tt http://cosmologist.info}~\cite{Lewis:2002ah}.

\ 

\subsection{Forecasts for CMB}

We produce forecasts on cosmological parameters for future CMB experiments with a well established and common method (see e.g. \cite{core1,Capparelli:2017tyx,Renzi:2017cbg}). Under the assumption of the fiducial model described previously, we compute the theoretical CMB angular spectra for temperature, $C_{\ell}^{TT}$, $E$ and $B$ modes polarization $C_{\ell}^{EE}$ and $C_{\ell}^{BB}$, and cross temperature-polarization $C_{\ell}^{TE}$, using the Boltzmann code ~\cite{camb}.

Given an experiment with FWHM angular resolution $\theta$ and experimental sensitivity $w^{-1}$ (expressed in $[\mu K$-arcmin$]^2$), we can introduce an  experimental noise for the temperature angular spectra of the form (see e.g. \cite{lesgourgues}): 

\begin{equation}
N_\ell = w^{-1}\exp(\ell(\ell+1)\theta^2/8\ln2).
\end{equation}

A similar expression is used to describe the noise for the polarization spectra with $w_p^{-1}=2w^{-1}$ (one detector measures two polarization states).

We have then produced synthetic realisations of CMB data assuming different possible future CMB experiments with technical specifications as listed in Table~\ref{tab:spec}. In particular, we have considered a possible future CMB satellite experiments such as LiteBIRD \cite{litebird} and three possible configurations for ground-based telescopes as Stage-III 'wide' (S3wide), Stage-III 'deep' (S3deep) (see \cite{erminia}), and CMB-S4 (see e.g. \cite{stage4,Capparelli:2017tyx,Renzi:2017cbg}).

The simulated experimental spectra are then compared with the theoretical spectra
using a likelihood ${\cal L}$ given by

\begin{equation}
 - 2 \ln {\cal L} = \sum_{l} (2l+1) f_{\rm sky} \left(
\frac{D}{|\bar{C}|} + \ln{\frac{|\bar{C}|}{|\hat{C}|}} - 3 \right),
\label{chieff}
	\end{equation}

\noindent where $\hat{C}_l$ are the theoretical spectra plus noise, while $\bar{C}_l$ are the fiducial spectra plus noise (i.e. our simulated dataset). The quantities $|\bar{C}|$, $|\hat{C}|$ are :

\begin{eqnarray}
|\bar{C}| &=& \bar{C}_\ell^{TT}\bar{C}_\ell^{EE}\bar{C}_\ell^{BB} -
\left(\bar{C}_\ell^{TE}\right)^2\bar{C}_\ell^{BB} ~, \\
|\hat{C}| &=& \hat{C}_\ell^{TT}\hat{C}_\ell^{EE}\hat{C}_\ell^{BB} -
\left(\hat{C}_\ell^{TE}\right)^2\hat{C}_\ell^{BB}~,
\end{eqnarray}

where $D$ is defined as
\begin{eqnarray}
D  &=&
\hat{C}_\ell^{TT}\bar{C}_\ell^{EE}\bar{C}_\ell^{BB} +
\bar{C}_\ell^{TT}\hat{C}_\ell^{EE}\bar{C}_\ell^{BB} +
\bar{C}_\ell^{TT}\bar{C}_\ell^{EE}\hat{C}_\ell^{BB} \nonumber\\
&&- \bar{C}_\ell^{TE}\left(\bar{C}_\ell^{TE}\hat{C}_\ell^{BB} +
2\hat{C}_\ell^{TE}\bar{C}_\ell^{BB} \right). \nonumber\\
\end{eqnarray}

In what follows we don't consider information from CMB lensing derived from trispectrum data.

\begin{table*}
\begin{center}
\begin{tabular}{lccccc}
\toprule
\horsp
Experiment \vertsp Beam \vertsp Power noise $w^{-1/2}$\vertsp $\ell_{max}$& $\ell_{min}$& $f_{sky}$\\
&  &[\footnotesize$\mu$K-arcmin]& & &\\
\hline
\morehorsp
LiteBIRD      & $30$' & $4.5$& $3000$&$2$&$0.7$\\
\morehorsp
S3deep      & $1$' & $4$ & $3000$&$50$&$0.06$\\
\morehorsp
S3wide      & $1.4$' & $8$ & $3000$&$50$&$0.4$\\
\morehorsp
CMB-S4      & $3$' & $1$& $3000$&$5,50$&$0.4$\\
\bottomrule
\end{tabular}
\end{center}
\caption{Specifications for the different experimental configurations considered in our paper. In case of polarization spectra the noise $w^{-1}$ is multiplied by a factor 2.}
\label{tab:spec}
\end{table*}

\
\subsection{Forecast for BAO}

For the future BAO dataset we consider the DESI experiment~\cite{Levi:2013gra}. If $D_V$ is the volume averaged distance, this is defined as:

\begin{equation}
D_V(z)\equiv \[\frac{\(1+z\)^2D_A(z)^2cz}{H(z)}\]^\frac{1}{3}
\end{equation}

\noindent where $D_A$ is the angular diameter distance and $H(z)$ the expansion rate.
Under the assumption of the fiducial model described previously, we compute the theoretical values of the ratio $r_s/D_V$, where $r_s$ is the sound horizon at the drag epoch when photons and baryons decouple, for the different redshifts in the range $z=[0.15-1.85]$ listed in Table~\ref{DESI}. Given the forecast uncertainties reported in~\cite{Font-Ribera:2013rwa} for $D_A/r_s$ and $H(z)$, we then compute the uncertainties on $r_s/D_V$ and we show them in Table~\ref{DESI}. The simulated BAO dataset is finally compared with the theoretical $r_s/D_V$ values through a Gaussian prior.

As a consistency test, we have checked that by using directly the $D_A/r_s$ value and the corresponding uncertainties reported in~\cite{Font-Ribera:2013rwa} instead of $r_s/D_V$, we obtain very similar results with constraints about $\sim 30 \%$ weaker on $H_0$ when combined with CMB-S4 data, in agreement with the results of \cite{addison}. 

In principle it would be possible to forecast BAO data considering $D_A/r_s$ and $H(z)$ as independent measurements. However some small tension (around $1$ sigma level) is present between the current constraints from $D_A/r_s$ and $H(z)$ (see e.g. \cite{addison}, Figure 2 contours in the Top Left and Bottom Left panels for $\Omega_m\sim0.3$).  It is clearly difficult to properly take into account a possible small tension between future $D_A/r_s$ and $H(z)$ measurements that could improve/reduce future BAO constraints. We therefore follow the approach of \cite{erminia} deriving the expected fractional uncertainties on $r_s/D_V$ for DESI from the fractional errors on $D_A/r_s$ and $H(z)$ forecasted in ~\cite{Font-Ribera:2013rwa}.

\begin{table}
\begin{center}
\begin{tabular}{lccc}
\toprule
\horsp
Redshift & $\frac{\sigma(r_s/D_V)}{r_s/D_V}$& $\sigma(r_s/D_V)$\\
\hline
\morehorsp
 $0.15$&$2.57\%$&$0.00595$\\
\morehorsp
 $0.25$&$1.71\%$&$0.00246$\\
\morehorsp
$0.35$&$1.32\%$&$0.00141$\\
\morehorsp
$0.45$&$1.08\%$&$0.00093$\\
\morehorsp
$0.55$&$0.91\%$&$0.00067$\\
\morehorsp
 $0.65$&$0.79\%$&$0.00051$\\
\morehorsp
 $0.75$&$0.70\%$&$0.00040$\\
\morehorsp
$0.85$&$0.68\%$&$0.00036$\\
\morehorsp
$0.95$&$0.75\%$&$0.00037$\\
\morehorsp
$1.05$&$0.77\%$&$0.00036$\\
\morehorsp
$1.15$&$0.76\%$&$0.00034$\\
\morehorsp
$1.25$&$0.76\%$&$0.00032$\\
\morehorsp
 $1.35$&$0.83\%$&$0.00033$\\
\morehorsp
 $1.45$&$0.96\%$&$0.00037$\\
\morehorsp
$1.55$&$1.21\%$&$0.00046$\\
\morehorsp
$1.65$&$1.89\%$&$0.00069$\\
\morehorsp
$1.75$&$2.91\%$&$0.00104$\\
\morehorsp
$1.85$&$3.87\%$&$0.00134$\\
\bottomrule
\end{tabular}
\end{center}
\caption{Specifications for the forecast DESI data, obtained by \cite{Font-Ribera:2013rwa}.}
\label{DESI}
\end{table}

\
\subsection{Forecast for standard sirens}

As stated in the introduction, in this paper we want to address the question of what kind of cosmological information can be obtained from GWSS systems within the coming decade (i.e. by $\sim2028$) when complementary measurements from CMB and BAO surveys will be available. We therefore focus our attention on GW experiments that could be completed in this time-scale: the Hanford-Livingston-Virgo (HLV) network of interferometers during the second year of operation at design sensitivity ($\sim2023$) and the the Hanford-Livingston-Virgo-Japan-India (HLVJI) network two years after the start of operations ($\sim 2026$)~\cite{2018LRR....21....3A}. We do not consider longer-term experiments such as the LISA \cite{lisa} or DECIGO \cite{decigo} missions or proposed third generation interferometers such as the Einstein Telescope \cite{ET} or the Cosmic Explore \cite{CE} that would presumably start operations no sooner than $2030$. Moreover, these experiments will be able to determine the luminosity distance of GWSS at higher redshift, opening the possibility to test the acceleration of the universe (i.e. the deceleration parameter), while here we only limit our discussion to the Hubble constant (although black holes standard sirens would probe these high redshifts earlier~\cite{2018arXiv180510270F}). 

Considering HLV or HLVJI and assuming the optimistic case that all binary neutron star (BNS) systems have detected optical counterparts and associated redshift measurements, the major uncertainty on the projected constraint on $H_0$ from GWSS comes from the BNS detection rate. The current best estimate of the BNS rate is $R=1540^{+3200}_{-1220}\,$Gpc$^{-3}$yr$^{-1}$~\cite{bnsrate} (median and $90 \%$ credible interval)~\cite{GW170817}; it is very poorly constrained given that only one BNS event has been detected to date. Following \cite{holz} we forecast $4 \%$, $2\%$, and $1 \%$ uncertainties on the measurement of $H_0$ for the HLV network after two years at design sensitivity ($\sim2023$) and assuming lower, mean, and upper BNS rates of $R= 320\,$Gpc$^{-3}$yr$^{-1}$, $R=1540\,$Gpc$^{-3}$yr$^{-1}$, and $R=4740\,$Gpc$^{-3}$yr$^{-1}$. The corresponding accuracy for the HLVJI network operating after one year of operation ($\sim 2025$) reaches $3 \%$, $1.4 \%$, and $0.8 \%$ on $H_0$, while after two years it arrives at $2.8 \%$, $1.2 \%$, and $0.7 \%$  (see Figure~$3$ in \cite{holz}). By 2028 the HLVJI network would have an additional two years of operation, leading {\em very roughly}\/ to a factor of $\sqrt{2}$ improvement to $2 \%$, $0.85 \%$, and $0.5 \%$.
It is therefore possible that standard siren measurements will reach an accuracy of $1 \%$ by $2028$ (under the assumption that a majority of BNS mergers have detectable electromagnetic counterparts). Considering that our fiducial model has $H_0=67.3$ km/s/Mpc, we therefore assume a Gaussian prior of $H_0=67.3 \pm 0.673$ km/s/Mpc.  In what follows we will refer to this (optimistic) prior as GWSS67.
On the other hand, we also consider the significantly more pessimistic $H_0$ prior of $4 \%$ ($H_0=67.3 \pm 2.7$ km/s/Mpc). This prior, just a factor of $\sim 4$ smaller than the current GW constraint based on a single event, is clearly extremely conservative but may happen if the BNS rate ends up on the low side (see e.g. \cite{GW170817, holz, rate1,rate2}). In what follows we will refer to this prior as PGWSS67.

These priors on $H_0$ are introduced by importance sampling on the models (samples) drawn from our MCMC simulations \cite{Lewis:2002ah}. In our case this translates into multiplying each sample weight by a Gaussian function, with mean and variance defined by the assumed $H_0$ prior, evaluated at the value of $H_0$ in the sample itself. For this to work it is only necessary that the obtained weights are significant for a large fraction of the re-weighted samples; this is a direct consequence of the requirement that the distribution from which the samples are drawn and the importance distribution are not too dissimilar.

\section{Results}

\subsection{$\Lambda$CDM+$\Omega_k$+$\Sigma m_{\nu}$+$N_{\rm eff}$+$w$ Model}

\begin{table*}[!hbtp]
\begin{center}
\begin{tabular}{lcccc}
\toprule
\horsp
Parameter \vertsp LiteBIRD \vertsp S3deep \vertsp S3wide \vertsp CMB-S4 \\
\hline\hline
\morehorsp
$\Omega_\mathrm{b}h^2$ \vertsp ${\siround{0.02214}{5}}\pm{\siround{0.00023}{5}}$\vertsp${\siround{0.02222}{5}}\pm{\siround{0.00016}{5}}$\vertsp${\siround{0.0222}{5}}\pm{\siround{9e-05}{5}}$\vertsp${\siround{0.02219}{5}}\pm{\siround{5e-05}{5}}$\\
\morehorsp
$\Omega_\mathrm{c}h^2$ \vertsp ${\siround{0.1203}{4}}\pm{\siround{0.0042}{4}}$\vertsp${\siround{0.1199}{4}}\pm{\siround{0.003}{4}}$\vertsp${\siround{0.1198}{4}}\pm{\siround{0.0013}{4}}$\vertsp${\siround{0.1199}{4}}\pm{\siround{0.001}{4}}$\\
\morehorsp
$100\theta_\mathrm{MC}$ \vertsp ${\siround{1.04075}{5}}\pm{\siround{0.00078}{5}}$\vertsp${\siround{1.04065}{5}}\pm{\siround{0.00033}{5}}$\vertsp${\siround{1.04071}{5}}\pm{\siround{0.00016}{5}}$\vertsp${\siround{1.04071}{5}}\pm{\siround{0.00012}{5}}$\\
\morehorsp
$\tau$ \vertsp ${\siround{0.054}{3}}\pm{\siround{0.002}{3}}$\vertsp${\siround{0.054}{3}}\pm{\siround{0.01}{3}}$\vertsp${\siround{0.053}{3}}\pm{\siround{0.01}{3}}$\vertsp${\siround{0.055}{3}}\pm{\siround{0.003}{3}}$\\
\morehorsp
$H_0$ \vertsp ${\siround{64}{2}}^{+\siround{8}{2}}_{-\siround{18}{2}}$\vertsp${\siround{59}{2}}^{+\siround{7}{2}}_{-\siround{19}{2}}$\vertsp${\siround{61}{2}}^{+\siround{7}{2}}_{-\siround{17}{2}}$\vertsp${\siround{60}{2}}^{+\siround{8}{2}}_{-\siround{11}{2}}$\\
\morehorsp
$\Omega_K$ \vertsp ${\siround{-0.014}{3}}^{+\siround{0.018}{3}}_{-\siround{0.005}{3}}$\vertsp${\siround{-0.027}{3}}^{+\siround{0.033}{3}}_{-\siround{0.01}{3}}$\vertsp${\siround{-0.016}{3}}^{+\siround{0.02}{3}}_{-\siround{0.006}{3}}$\vertsp${\siround{-0.012}{3}}^{+\siround{0.016}{3}}_{-\siround{0.004}{3}}$\\
\morehorsp
$\log(10^{10} A_\mathrm{s})$ \vertsp ${\siround{3.092}{3}}\pm{\siround{0.011}{3}}$\vertsp${\siround{3.09}{3}}\pm{\siround{0.021}{3}}$\vertsp${\siround{3.09}{3}}\pm{\siround{0.021}{3}}$\vertsp${\siround{3.093}{3}}\pm{\siround{0.006}{3}}$\\
\morehorsp
$n_\mathrm{s}$ \vertsp ${\siround{0.9629}{4}}^{+\siround{0.0073}{4}}_{-\siround{0.0074}{4}}$\vertsp${\siround{0.9656}{4}}\pm{\siround{0.0112}{4}}$\vertsp${\siround{0.965}{4}}^{+\siround{0.0048}{4}}_{-\siround{0.0046}{4}}$\vertsp${\siround{0.9648}{4}}\pm{\siround{0.0038}{4}}$\\
\morehorsp
$w$ \vertsp ${\siround{-1.069}{3}}^{+\siround{0.638}{3}}_{-\siround{0.297}{3}}$\vertsp${\siround{-0.896}{3}}^{+\siround{0.661}{3}}_{-\siround{0.279}{3}}$\vertsp${\siround{-0.911}{3}}^{+\siround{0.506}{3}}_{-\siround{0.243}{3}}$\vertsp${\siround{-0.846}{3}}^{+\siround{0.283}{3}}_{-\siround{0.234}{3}}$\\
\morehorsp
$N_{\rm eff}$ \vertsp ${\siround{3.069}{3}}^{+\siround{0.243}{3}}_{-\siround{0.246}{3}}$\vertsp${\siround{3.082}{3}}\pm{\siround{0.141}{3}}$\vertsp${\siround{3.063}{3}}\pm{\siround{0.07}{3}}$\vertsp${\siround{3.06}{3}}^{+\siround{0.046}{3}}_{-\siround{0.045}{3}}$\\
\morehorsp
$\Sigma m_\nu$\vertsp $< \siround{0.594}{3}$ eV\vertsp$< \siround{0.584}{3}$ eV\vertsp$< \siround{0.405}{3}$ eV\vertsp$< \siround{0.322}{3}$ eV\\\bottomrule
\end{tabular}
\caption{Forecasted constraints at $68 \%$ C.L. (upper limits at $95 \%$ C.L.) from future CMB experiments with specifications listed in Table~\ref{tab:spec} in an extended $\Lambda$CDM+$\Omega_k$+$\Sigma m_{\nu}$+$N_{\rm eff}$+$w$ $10$ parameters analysis. A $6$ parameters $\Lambda$CDM model is assumed as fiducial model. Parameters as $H_0$ and $w$ are practically unbounded. $\Omega_k$ and $\Sigma m_{\nu}$ are also weakly constrained.}
\label{CMBconstraints}
\end{center}
\end{table*}
\vspace{.1cm}
\begin{figure*}[!hbtp]
\includegraphics[width=.67\textwidth,keepaspectratio]{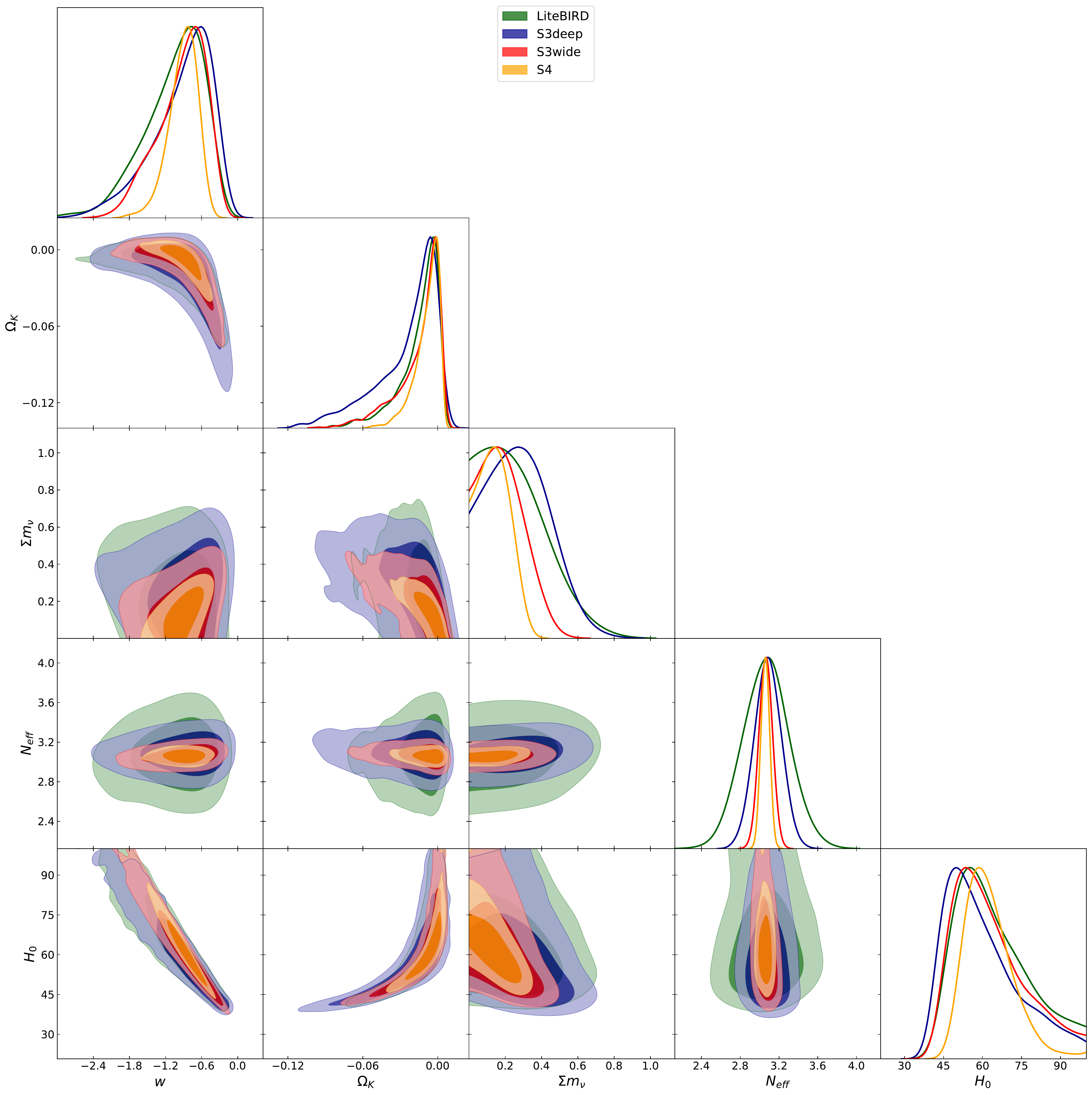}
\caption{Forecasted future constraints at $68 \%$ and $95 \%$ C.L. from future CMB data for the experimental configurations in Table~\ref{tab:spec} in case of the $\Lambda$CDM+$\Omega_k$+$\Sigma m_{\nu}$+$N_{\rm eff}$+$w$ extended model. Clearly in this extended parameter space CMB data alone will be unable to significantly constrain geometrical parameters as $H_0$, $\Omega_k$ or $w$.}
\label{cmbalone}
\end{figure*}

We first forecast the constraints on cosmological parameters from future CMB data only, assuming the extended $10$ parameter model $\Lambda$CDM+$\Omega_k$+$\Sigma m_{\nu}$+$N_{\rm eff}$+$w$. The constraints on cosmological parameters for the experimental configurations listed in Table~\ref{tab:spec} are reported in Table~\ref{CMBconstraints}, while 2D contour plots at $68 \%$ C.L. and $95 \%$ C.L. between the extra parameters are reported in Figure~\ref{cmbalone}.
We find that future experiments, including CMB-S4, will be unable to provide significant additional constraints on geometrical parameters such as $H_0$, $\Omega_k$, and $w$. This is due to the well known geometrical degeneracy that affects CMB observables (see, e.g.,~\cite{deg1,deg2,deg3}).
CMB-S4 will improve the constraints on $n_S$, $N_{\rm eff}$, $\Omega_bh^2$,  and $\Omega_ch^2$ by a factor of $\sim 2$--5 with respect to LiteBIRD. These parameters are less affected by the geometrical degeneracy, and can thus be better constrained with an improvement in the angular resolution of the experiment. Constraints on neutrino masses will also only see marginal improvement (i.e. $\Sigma m_{\nu}<0.32$ eV at $95 \%$ C.L. for the strongest case from CMB-S4), which falls short of the sensitivity of $\Delta \Sigma m_{\nu} \sim 0.05\,$eV needed to test the inverted neutrino mass hierarchy at two standard deviations. The neutrino effective number will be, on the contrary, less affected and interesting constraints at the $\Delta N_{\rm eff} \sim 0.045$ level can be achieved with CMB-S4 even in the case of a very extended parameter space.

\begin{table*}[!hbtp]
\begin{center}
\begin{tabular}{lcccc}
\toprule
\horsp
Parameter \vertsp LiteBIRD+DESI \vertsp S3deep+DESI \vertsp S3wide+DESI \vertsp CMB-S4+DESI \\
\hline\hline
\morehorsp
$\Omega_\mathrm{b}h^2$ \vertsp ${\siround{0.02219}{5}}\pm{\siround{0.00022}{5}}$\vertsp${\siround{0.02219}{5}}\pm{\siround{0.00016}{5}}$\vertsp${\siround{0.02218}{5}}\pm{\siround{9e-05}{5}}$\vertsp${\siround{0.02218}{5}}\pm{\siround{5e-05}{5}}$\\
\morehorsp
$\Omega_\mathrm{c}h^2$ \vertsp ${\siround{0.1212}{4}}^{+\siround{0.0033}{4}}_{-\siround{0.0041}{4}}$\vertsp${\siround{0.1208}{4}}\pm{\siround{0.0027}{4}}$\vertsp${\siround{0.1199}{4}}\pm{\siround{0.0013}{4}}$\vertsp${\siround{0.1199}{4}}\pm{\siround{0.001}{4}}$\\
\morehorsp
$100\theta_\mathrm{MC}$ \vertsp ${\siround{1.04058}{5}}^{+\siround{0.00071}{5}}_{-\siround{0.0007}{5}}$\vertsp${\siround{1.04069}{5}}\pm{\siround{0.00031}{5}}$\vertsp${\siround{1.04075}{5}}\pm{\siround{0.00015}{5}}$\vertsp${\siround{1.04076}{5}}\pm{\siround{0.00011}{5}}$\\
\morehorsp
$\tau$ \vertsp ${\siround{0.055}{3}}\pm{\siround{0.002}{3}}$\vertsp${\siround{0.057}{3}}\pm{\siround{0.009}{3}}$\vertsp${\siround{0.057}{3}}\pm{\siround{0.008}{3}}$\vertsp${\siround{0.055}{3}}^{+\siround{0.002}{3}}_{-\siround{0.003}{3}}$\\
\morehorsp
$H_0$ \vertsp ${\siround{67.8}{1}}^{+\siround{1.3}{1}}_{-\siround{1.5}{1}}$\vertsp${\siround{67.7}{1}}^{+\siround{1.2}{1}}_{-\siround{1.3}{1}}$\vertsp${\siround{67.4}{1}}^{+\siround{1.0}{1}}_{-\siround{1.2}{1}}$\vertsp${\siround{67.4}{1}}^{+\siround{1.0}{1}}_{-\siround{1.1}{1}}$\\
\morehorsp
$\Omega_K$ \vertsp ${\siround{0.0}{3}}^{+\siround{0.001}{3}}_{-\siround{0.002}{3}}$\vertsp${\siround{0.001}{3}}\pm{\siround{0.002}{3}}$\vertsp${\siround{0.0}{3}}\pm{\siround{0.001}{3}}$\vertsp${\siround{0.0}{3}}\pm{\siround{0.001}{3}}$\\
\morehorsp
$\log(10^{10} A_\mathrm{s})$ \vertsp ${\siround{3.097}{3}}\pm{\siround{0.009}{3}}$\vertsp${\siround{3.101}{3}}\pm{\siround{0.018}{3}}$\vertsp${\siround{3.099}{3}}\pm{\siround{0.016}{3}}$\vertsp${\siround{3.095}{3}}^{+\siround{0.005}{3}}_{-\siround{0.006}{3}}$\\
\morehorsp
$n_\mathrm{s}$ \vertsp ${\siround{0.9656}{4}}^{+\siround{0.0069}{4}}_{-\siround{0.0068}{4}}$\vertsp${\siround{0.9637}{4}}\pm{\siround{0.0104}{4}}$\vertsp${\siround{0.9645}{4}}^{+\siround{0.0046}{4}}_{-\siround{0.0047}{4}}$\vertsp${\siround{0.9647}{4}}^{+\siround{0.0037}{4}}_{-\siround{0.0036}{4}}$\\
\morehorsp
$w$ \vertsp ${\siround{-1.013}{3}}^{+\siround{0.054}{3}}_{-\siround{0.047}{3}}$\vertsp${\siround{-1.022}{3}}^{+\siround{0.057}{3}}_{-\siround{0.047}{3}}$\vertsp${\siround{-1.01}{3}}^{+\siround{0.051}{3}}_{-\siround{0.045}{3}}$\vertsp${\siround{-1.005}{3}}^{+\siround{0.047}{3}}_{-\siround{0.043}{3}}$\\
\morehorsp
$N_{\rm eff}$ \vertsp ${\siround{3.118}{3}}^{+\siround{0.206}{3}}_{-\siround{0.237}{3}}$\vertsp${\siround{3.065}{3}}^{+\siround{0.136}{3}}_{-\siround{0.138}{3}}$\vertsp${\siround{3.049}{3}}\pm{\siround{0.067}{3}}$\vertsp${\siround{3.051}{3}}\pm{\siround{0.044}{3}}$\\
\morehorsp
$\Sigma m_\nu$\vertsp $< \siround{0.202}{3}$ eV\vertsp$< \siround{0.253}{3}$ eV\vertsp$< \siround{0.186}{3}$ eV\vertsp$< \siround{0.126}{3}$ eV\\
\bottomrule
\end{tabular}
\caption{Forecasted constraints at $68 \%$ C.L. (upper limits at $95 \%$ C.L.) from future CMB experiments with specifications listed in Table~\ref{tab:spec} plus information from the BAO DESI galaxy survey in an extended $\Lambda$CDM+$\Omega_k$+$\Sigma m_{\nu}$+$N_{\rm eff}$+$w$, $10$ parameters, analysis. A $6$ parameters $\Lambda$CDM model is assumed as fiducial model. When comparing the results with those in the CMB alone case reported in Table~\ref{CMBconstraints} we can notice a significant improvement in geometrical parameters as $H_0$, $w$ and $\Omega_k$. Constraints on neutrino masses are also improved.}
\label{CMBDESIconstraints}
\end{center}
\end{table*}

\begin{figure*}[!hbtp]
\includegraphics[width=.67\textwidth,keepaspectratio]{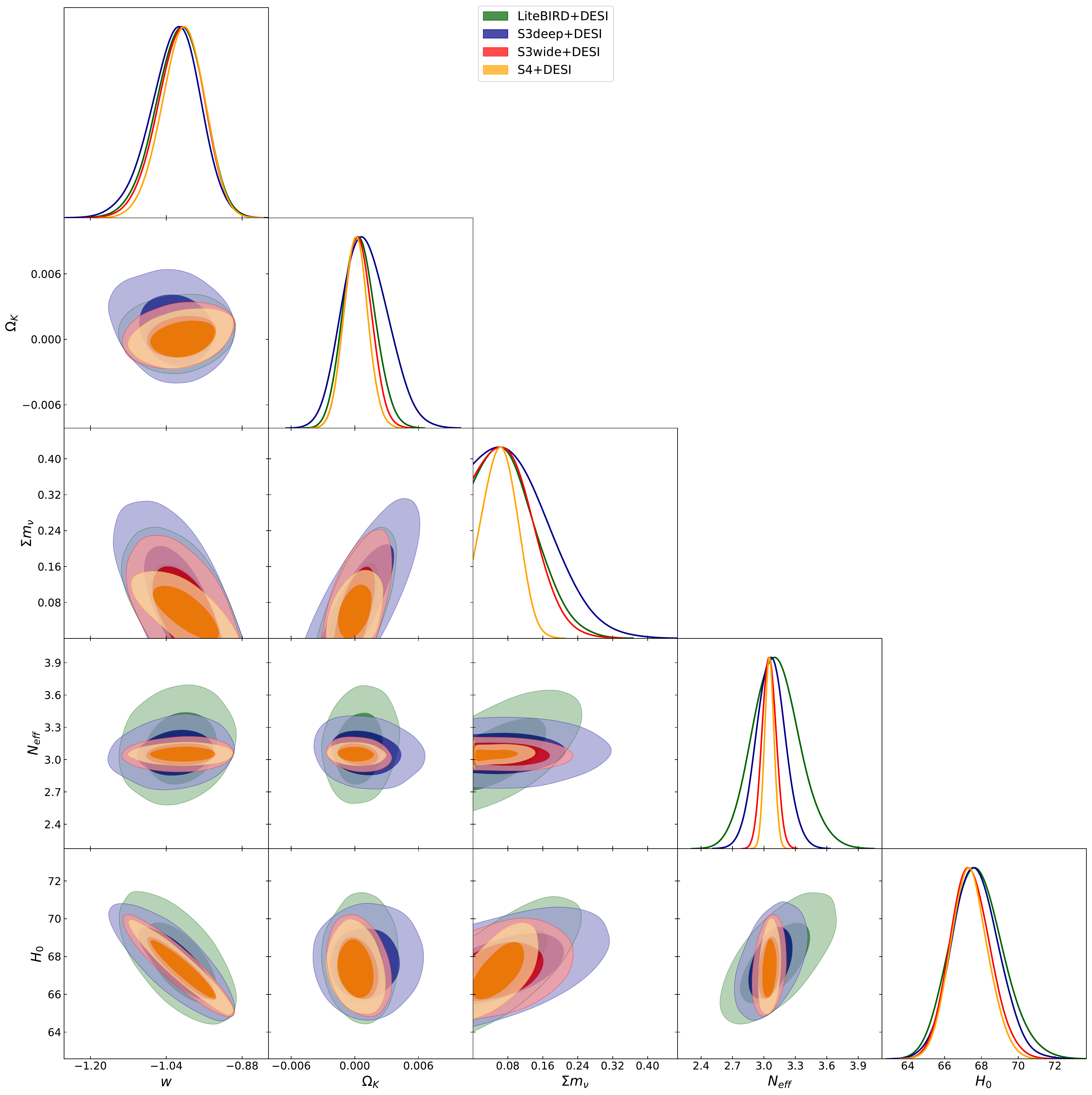}
\caption{Forecasted constraints at $68 \%$ and $95 \%$ C.L.  from CMB+DESI data for the experimental configurations in Table~\ref{tab:spec} in case of the $\Lambda$CDM+$\Omega_k$+$\Sigma m_{\nu}$+$N_{\rm eff}$+$w$ extended model.}
\label{cmbdesi}
\end{figure*}

It is interesting to investigate how the inclusion of future BAO surveys, such as DESI, can break the geometrical degeneracy and improve the constraints derived from CMB data. Assuming the same $\Lambda$CDM fiducial model, we report the CMB+DESI constraints in  Table~\ref{CMBDESIconstraints} and we show the 2D confidence level contours at $68 \%$ C.L. and $95 \%$ C.L. in Figure~\ref{cmbdesi}.
The geometrical parameters are constrained almost equally by all configurations, indicating that the additional constraining power arises from the inclusion of DESI. Curvature is now determined with a 0.1--0.2\% accuracy, while the equation of state can be determined with a $\sim 5 \%$ accuracy. It is interesting to note that a degeneracy is present between $\Omega_k$, $w$, and $\Sigma m_{\nu}$, i.e. the introduction of a neutrino mass limits the CMB+BAO constraints on curvature and $w$. In addition, after the inclusion of DESI, CMB-S4+DESI provides better constraints by a factor $\sim 2-4$ on  parameters such as $n_S$ and $N_{\rm eff}$ with respect to LiteBIRD+DESI. The bounds on the sum of neutrino masses are however still affected by the remaining extra parameters (mostly by the anti-correlation with $w$ and the correlation with $\Omega_k$), resulting in a limit of $\Sigma m_{\nu} <0.126$ eV at $95 \%$ C.L. for the CMB-S4+DESI configuration and $\Sigma m_{\nu} <0.202$ eV at $95 \%$ C.L. for LiteBIRD+DESI. However the key result for our analysis is the constraint on the Hubble parameter. Again, between the several configurations we consider, CMB-S4+DESI provides the best constraint of $H_0=67.4^{+1.0}_{-1.1}\,$km/s/Mpc, i.e. an uncertainty on the value of the Hubble constant of the order of $\sim 1.5 \%$, while LiteBIRD+DESI gives $H_0=67.8^{+1.3}_{-1.5}\,$km/s/Mpc with an uncertainty of $\sim 2 \%$.

\begin{table*}[!hbtp]
\begin{center}
\begin{tabular}{lcccc}
\toprule
\horsp
Parameter \vertsp LiteBIRD+GWSS67 \vertsp S3deep+GWSS67 \vertsp S3wide+GWSS67 \vertsp CMB-S4+GWSS67 \\
\hline\hline
\morehorsp
$\Omega_\mathrm{b}h^2$ \vertsp ${\siround{0.02215}{5}}\pm{\siround{0.00023}{5}}$\vertsp${\siround{0.02221}{5}}\pm{\siround{0.00017}{5}}$\vertsp${\siround{0.0222}{5}}\pm{\siround{9e-05}{5}}$\vertsp${\siround{0.02219}{5}}\pm{\siround{5e-05}{5}}$\\
\morehorsp
$\Omega_\mathrm{c}h^2$ \vertsp ${\siround{0.1204}{4}}^{+\siround{0.0042}{4}}_{-\siround{0.0043}{4}}$\vertsp${\siround{0.1199}{4}}^{+\siround{0.0032}{4}}_{-\siround{0.003}{4}}$\vertsp${\siround{0.1198}{4}}^{+\siround{0.0014}{4}}_{-\siround{0.0013}{4}}$\vertsp${\siround{0.12}{4}}^{+\siround{0.001}{4}}_{-\siround{0.0009}{4}}$\\
\morehorsp
$100\theta_\mathrm{MC}$ \vertsp ${\siround{1.04075}{5}}\pm{\siround{0.0008}{5}}$\vertsp${\siround{1.04068}{5}}^{+\siround{0.00031}{5}}_{-\siround{0.00035}{5}}$\vertsp${\siround{1.04074}{5}}^{+\siround{0.00015}{5}}_{-\siround{0.00016}{5}}$\vertsp${\siround{1.04075}{5}}\pm{\siround{0.00011}{5}}$\\
\morehorsp
$\tau$ \vertsp ${\siround{0.055}{3}}\pm{\siround{0.002}{3}}$\vertsp${\siround{0.054}{3}}\pm{\siround{0.01}{3}}$\vertsp${\siround{0.053}{3}}\pm{\siround{0.011}{3}}$\vertsp${\siround{0.055}{3}}^{+\siround{0.002}{3}}_{-\siround{0.003}{3}}$\\
\morehorsp
$H_0$ \vertsp ${\siround{67.3}{2}}^{+\siround{0.67}{2}}_{-\siround{0.68}{2}}$\vertsp${\siround{67.3}{2}}^{+\siround{0.65}{2}}_{-\siround{0.67}{2}}$\vertsp${\siround{67.26}{2}}^{+\siround{0.66}{2}}_{-\siround{0.63}{2}}$\vertsp${\siround{67.27}{2}}\pm{\siround{0.65}{2}}$\\
\morehorsp
$\Omega_K$ \vertsp ${\siround{-0.005}{3}}^{+\siround{0.007}{3}}_{-\siround{0.005}{3}}$\vertsp${\siround{-0.006}{3}}^{+\siround{0.007}{3}}_{-\siround{0.008}{3}}$\vertsp${\siround{-0.004}{3}}\pm{\siround{0.005}{3}}$\vertsp${\siround{-0.001}{3}}\pm{\siround{0.003}{3}}$\\
\morehorsp
$\log(10^{10} A_\mathrm{s})$ \vertsp ${\siround{3.093}{3}}\pm{\siround{0.01}{3}}$\vertsp${\siround{3.091}{3}}^{+\siround{0.022}{3}}_{-\siround{0.023}{3}}$\vertsp${\siround{3.09}{3}}\pm{\siround{0.022}{3}}$\vertsp${\siround{3.095}{3}}^{+\siround{0.005}{3}}_{-\siround{0.006}{3}}$\\
\morehorsp
$n_\mathrm{s}$ \vertsp ${\siround{0.9631}{4}}^{+\siround{0.0072}{4}}_{-\siround{0.0074}{4}}$\vertsp${\siround{0.9658}{4}}^{+\siround{0.0117}{4}}_{-\siround{0.0104}{4}}$\vertsp${\siround{0.9653}{4}}^{+\siround{0.0049}{4}}_{-\siround{0.0047}{4}}$\vertsp${\siround{0.9649}{4}}^{+\siround{0.0035}{4}}_{-\siround{0.0037}{4}}$\\
\morehorsp
$w$ \vertsp ${\siround{-1.199}{3}}^{+\siround{0.26}{3}}_{-\siround{0.112}{3}}$\vertsp${\siround{-1.208}{3}}^{+\siround{0.241}{3}}_{-\siround{0.142}{3}}$\vertsp${\siround{-1.1}{3}}^{+\siround{0.126}{3}}_{-\siround{0.086}{3}}$\vertsp${\siround{-1.032}{3}}^{+\siround{0.07}{3}}_{-\siround{0.046}{3}}$\\
\morehorsp
$N_{\rm eff}$ \vertsp ${\siround{3.073}{3}}^{+\siround{0.243}{3}}_{-\siround{0.255}{3}}$\vertsp${\siround{3.076}{3}}^{+\siround{0.147}{3}}_{-\siround{0.141}{3}}$\vertsp${\siround{3.059}{3}}\pm{\siround{0.07}{3}}$\vertsp${\siround{3.055}{3}}^{+\siround{0.044}{3}}_{-\siround{0.043}{3}}$\\
\morehorsp
$\Sigma m_\nu$\vertsp $< \siround{0.587}{3}$ eV\vertsp$< \siround{0.536}{3}$ eV\vertsp$< \siround{0.326}{3}$ eV\vertsp$< \siround{0.206}{3}$ eV\\\bottomrule
\end{tabular}
\caption{Forecasted constraints at $68 \%$ C.L. (upper limits at $95 \%$ C.L.)  from CMB+GWSS67 data for the experimental configurations in Table~\ref{tab:spec} in case of the $\Lambda$CDM+$\Omega_k$+$\Sigma m_{\nu}$+$N_{\rm eff}$+$w$ extended model.}
\label{CMBGW67constraints}
\end{center}
\end{table*}

\begin{figure*}[!hbtp]
\includegraphics[width=.67\textwidth,keepaspectratio]{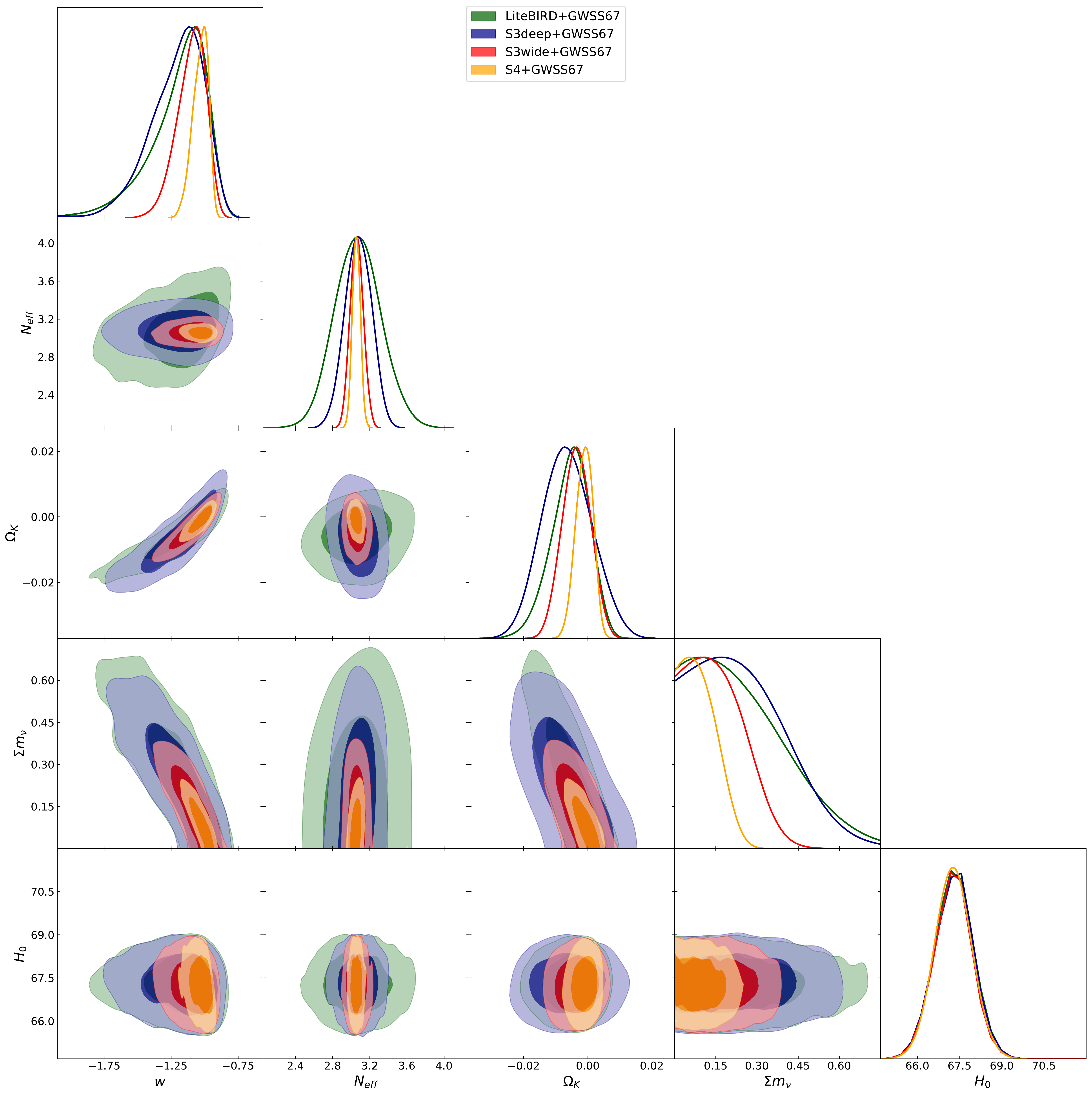}
\caption{Forecasted constraints at $68 \%$ and $95 \%$ C.L.  from CMB+GWSS67 data for the experimental configurations in Table~\ref{tab:spec} in case of the $\Lambda$CDM+$\Omega_k$+$\Sigma m_{\nu}$+$N_{\rm eff}$+$w$ extended model.}
\label{cmbgw67}
\end{figure*}

As discussed in the previous section, a similar uncertainty can be reached by the HLVJI network after one year of observations ($\sim2025$)  with a BNS detection rate of $R\ge 1540$ Gpc$^{-3}$yr$^{-1}$ or by HLV after two years of observations($\sim2023$) if the rate is $R\ge2800$ Gpc$^{-3}$yr$^{-1}$. For simplicity we have assumed that the standard siren accuracy on $H_0$ scales as $1/\sqrt{N_{\rm BNS}}$ where $N_{\rm BNS}$ is the number of observed BNS systems, which is a good approximation for $N\gtrsim20$~\cite{holz}. A first conclusion is that by 2025--2030 standard sirens may offer constraints on $H_0$ that are comparable in accuracy to those achievable from future CMB+BAO missions at a similar epoch.

\begin{figure*}[!hbtp]
\includegraphics[width=.67\textwidth,keepaspectratio]{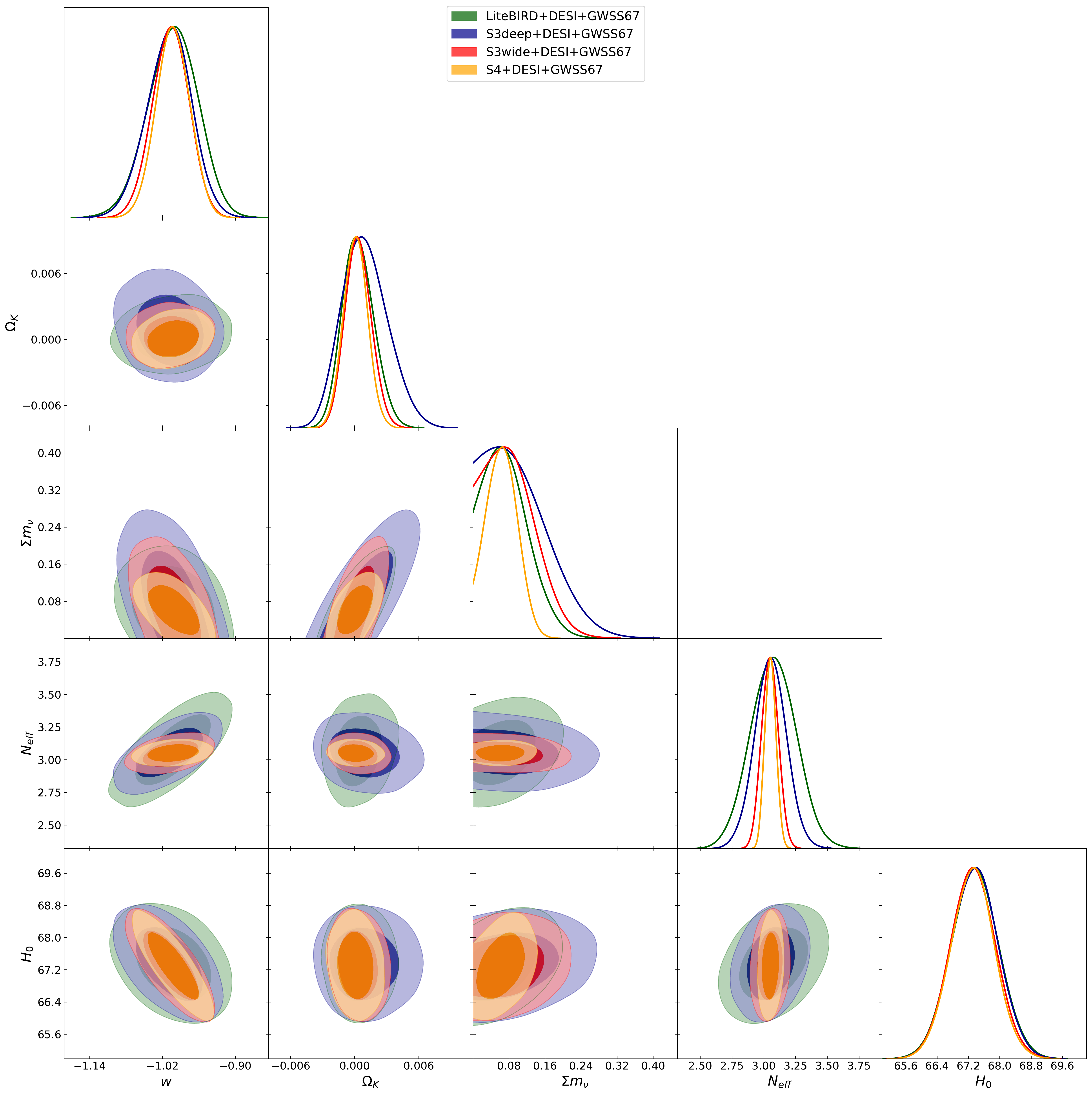}
\caption{Forecasted constraints at $68 \%$ and $95 \%$ C.L.  from CMB+DESI+GWSS67 data for the experimental configurations in Table~\ref{tab:spec} in case of the $\Lambda$CDM+$\Omega_k$+$\Sigma m_{\nu}$+$N_{\rm eff}$+$w$ extended model.}
\label{cmbdesigw670}
\end{figure*}

Furthermore, given existing estimates of the BNS event rate, an even higher $H_0$ accuracy may be expected from GWSS.
In Table~\ref{CMBGW67constraints} and in Figure~\ref{cmbgw67} we report the future constraints achievable by a combination of the CMB data and a prior on the Hubble constant with a $1 \%$ accuracy (GWSS67).  This GWSS67 prior, with respect to the CMB data alone, breaks the geometrical degeneracy and improves significantly the constraints on the corresponding parameters, now producing strong bounds on cosmological parameters such as curvature ($0.3 \%$ accuracy from CMB-S4+GWSS67) and $w$ ($7 \%$ accuracy from CMB-S4+GWSS67). The bound on neutrino masses is improved by $\sim 30 \%$, while there is no significant improvement on the remaining parameters ($N_{\rm eff}$, $n_S$, and the cold dark matter and baryon densities).
How would the inclusion of a GWSS measurement of $H_0$ impact cosmological constraints derived from a CMB+DESI? We answer to this question in Table~\ref{CMBDESIGW67constraints} and Figure~\ref{cmbdesigw670} where we report the constraints achievable from the full combined dataset.
We find that the combined analysis (in the case of CMB-S4) would constrain the Hubble constant with an accuracy of $\sim 0.5\,$km/s/Mpc, i.e. nearly a factor of two better than the CMB-S4+DESI case. A similar improvement is present with respect to LiteBIRD+DESI. Constraints on the dark energy equation of state are also significantly improved, by 30--40\%, reaching an accuracy of about $3\%$ with CMB-S4+DESI and $4 \%$ with LiteBIRD+DESI.

It is interesting to note that the constraints on $H_0$, $\Omega_k$, and $w$ coming from a combined analysis of DESI, GWSS, and a CMB mission such as LiteBIRD, S3deep, or S3wide, will be comparable or in some cases even better than the corresponding constraints coming from a CMB-S4+DESI dataset. For example, a $0.1\%$ accuracy on curvature or a $3 \%$ accuracy on $w$ can be reached by a S3wide+DESI+GWSS67 configuration instead of CMB-S4+DESI.
Alternatively, the GWSS measurement would also provide an interesting consistency check between different CMB+BAO datasets. 

\begin{table*}[!hbtp]
\begin{center}
\begin{tabular}{lcccc}
\toprule
\horsp
Parameter \vertsp LiteBIRD+DESI+GWSS67 \vertsp S3deep+DESI+GWSS67 \vertsp S3wide+DESI+GWSS67 \vertsp CMB-S4+DESI+GWSS67\\
\hline\hline
\morehorsp
$\Omega_\mathrm{b}h^2$ \vertsp ${\siround{0.02218}{5}}\pm{\siround{0.00021}{5}}$\vertsp${\siround{0.02218}{5}}^{+\siround{0.00015}{5}}_{-\siround{0.00016}{5}}$\vertsp${\siround{0.02218}{5}}\pm{\siround{9e-05}{5}}$\vertsp${\siround{0.02218}{5}}\pm{\siround{5e-05}{5}}$\\
\morehorsp
$\Omega_\mathrm{c}h^2$ \vertsp ${\siround{0.1205}{4}}^{+\siround{0.0028}{4}}_{-\siround{0.0031}{4}}$\vertsp${\siround{0.1205}{4}}\pm{\siround{0.0024}{4}}$\vertsp${\siround{0.1199}{4}}\pm{\siround{0.0013}{4}}$\vertsp${\siround{0.1199}{4}}\pm{\siround{0.001}{4}}$\\
\morehorsp
$100\theta_\mathrm{MC}$ \vertsp ${\siround{1.04069}{5}}^{+\siround{0.00064}{5}}_{-\siround{0.00063}{5}}$\vertsp${\siround{1.04072}{5}}\pm{\siround{0.0003}{5}}$\vertsp${\siround{1.04075}{5}}\pm{\siround{0.00015}{5}}$\vertsp${\siround{1.04076}{5}}\pm{\siround{0.00011}{5}}$\\
\morehorsp
$\tau$ \vertsp ${\siround{0.055}{3}}\pm{\siround{0.002}{3}}$\vertsp${\siround{0.057}{3}}\pm{\siround{0.009}{3}}$\vertsp${\siround{0.057}{3}}\pm{\siround{0.008}{3}}$\vertsp${\siround{0.055}{3}}^{+\siround{0.002}{3}}_{-\siround{0.003}{3}}$\\
\morehorsp
$H_0$ \vertsp ${\siround{67.37}{2}}^{+\siround{0.6}{2}}_{-\siround{0.61}{2}}$\vertsp${\siround{67.36}{2}}^{+\siround{0.58}{2}}_{-\siround{0.59}{2}}$\vertsp${\siround{67.32}{2}}\pm{\siround{0.57}{2}}$\vertsp${\siround{67.31}{2}}^{+\siround{0.54}{2}}_{-\siround{0.55}{2}}$\\
\morehorsp
$\Omega_K$ \vertsp ${\siround{0.0}{3}}^{+\siround{0.001}{3}}_{-\siround{0.002}{3}}$\vertsp${\siround{0.001}{3}}\pm{\siround{0.002}{3}}$\vertsp${\siround{0.0}{3}}\pm{\siround{0.001}{3}}$\vertsp${\siround{0.0}{3}}\pm{\siround{0.001}{3}}$\\
\morehorsp
$\log(10^{10} A_\mathrm{s})$ \vertsp ${\siround{3.096}{3}}\pm{\siround{0.008}{3}}$\vertsp${\siround{3.1}{3}}\pm{\siround{0.018}{3}}$\vertsp${\siround{3.099}{3}}\pm{\siround{0.016}{3}}$\vertsp${\siround{3.095}{3}}^{+\siround{0.005}{3}}_{-\siround{0.006}{3}}$\\
\morehorsp
$n_\mathrm{s}$ \vertsp ${\siround{0.9648}{4}}\pm{\siround{0.0061}{4}}$\vertsp${\siround{0.9635}{4}}\pm{\siround{0.01}{4}}$\vertsp${\siround{0.9645}{4}}\pm{\siround{0.0046}{4}}$\vertsp${\siround{0.9648}{4}}\pm{\siround{0.0036}{4}}$\\
\morehorsp
$w$ \vertsp ${\siround{-1.003}{3}}^{+\siround{0.043}{3}}_{-\siround{0.039}{3}}$\vertsp${\siround{-1.009}{3}}^{+\siround{0.038}{3}}_{-\siround{0.035}{3}}$\vertsp${\siround{-1.007}{3}}\pm{\siround{0.03}{3}}$\vertsp${\siround{-1.003}{3}}\pm{\siround{0.028}{3}}$\\
\morehorsp
$N_{\rm eff}$ \vertsp ${\siround{3.076}{3}}^{+\siround{0.176}{3}}_{-\siround{0.178}{3}}$\vertsp${\siround{3.053}{3}}^{+\siround{0.124}{3}}_{-\siround{0.123}{3}}$\vertsp${\siround{3.048}{3}}^{+\siround{0.065}{3}}_{-\siround{0.066}{3}}$\vertsp${\siround{3.052}{3}}^{+\siround{0.043}{3}}_{-\siround{0.044}{3}}$\\
\morehorsp
$\Sigma m_\nu$\vertsp $< \siround{0.164}{3}$ eV\vertsp$< \siround{0.226}{3}$ eV\vertsp$< \siround{0.18}{3}$ eV\vertsp$< \siround{0.12}{3}$ eV\\\bottomrule
\end{tabular}
\caption{Forecasted constraints at $68 \%$ C.L. (upper limits at $95 \%$ C.L.) from CMB+DESI+GWSS67 data for the experimental configurations in Table~\ref{tab:spec} in case of the $\Lambda$CDM+$\Omega_k$+$\Sigma m_{\nu}$+$N_{\rm eff}$+$w$ extended model.}
\label{CMBDESIGW67constraints}
\end{center}
\end{table*}

We also consider the possibility that future standard siren measurements of $H_0$ will confirm the current tension on the Hubble constant between CMB+BAO and local measurements from supernovae. It is interesting to evaluate at how many standard deviations a CMB+DESI measurement of $H_0$ will disagree with a GWSS determination of $H_0=73.30\pm0.73\,$km/s/Mpc.
From Table~\ref{CMBDESIconstraints}, we find that the standard siren measurement would be $4$ standard deviations from the expected LiteBIRD+DESI constraint, and at roughly $5$ standard deviations from the CMB-S4+DESI value. This is a significant improvement, since in an extended parameter space such the one we are considering the existing tension is at about $2$ standard deviations (see e.g. \cite{papero}).

Finally, let us consider a significantly more pessimistic GW prior on $H_0$ with a $\sim 4\%$ accuracy (PGWSS67).
In Table~\ref{CMBPGW67constraints} we report the constraints achievable from a combination of this prior with future CMB data. As expected, the constraints on curvature and $w$ are relaxed respect to the previous analyses of CMB+GWSS67 but only by a $\sim 10-20 \%$. In practice, the geometrical degeneracies between cosmological parameters present in CMB data only can be already sufficiently broken with a, pessimistic, PGWSS67 prior. An improvement of a factor four in the determination of $H_0$ will result in a, more modest, $10 \%$ improvement in the parameters. A first conclusion is therefore that in this theoretical framework, the GWSS67 and the PGWSS67 prior produce very similar constraints when combined with CMB data. 
On the other hand, combining the PGWSS67 prior with CMB+DESI data has a small effect in improving the constraints on $w$. We have found that in this case the constraints on $w$ improve just by $\sim 5 \%$ while, as discussed above, the improvement in case of GWSS67 is larger than $\sim 20 \%$.
The $4 \%$ PGWSS67 prior will clearly provide little help in solving the current tension on the value of the Hubble parameter.

\begin{table*}[!hbtp]
\begin{center}
\begin{tabular}{lcccc}
\toprule
\horsp
Parameter \vertsp LiteBIRD+PGWSS67($4\%$) \vertsp S3deep+PGWSS67($4\%$) \vertsp S3wide+PGWSS67($4\%$) \vertsp S4+PGWSS67($4\%$) \\
\hline\hline
\horsp
$\Omega_\mathrm{b}h^2$ \vertsp ${\siround{0.02215}{5}}\pm{\siround{0.00023}{5}}$\vertsp${\siround{0.02222}{5}}\pm{\siround{0.00017}{5}}$\vertsp${\siround{0.02219}{5}}\pm{\siround{9e-05}{5}}$\vertsp${\siround{0.02219}{5}}\pm{\siround{5e-05}{5}}$\\
\morehorsp
$\Omega_\mathrm{c}h^2$ \vertsp ${\siround{0.1204}{4}}^{+\siround{0.004}{4}}_{-\siround{0.0044}{4}}$\vertsp${\siround{0.1199}{4}}^{+\siround{0.0031}{4}}_{-\siround{0.0029}{4}}$\vertsp${\siround{0.1198}{4}}^{+\siround{0.0014}{4}}_{-\siround{0.0013}{4}}$\vertsp${\siround{0.12}{4}}^{+\siround{0.001}{4}}_{-\siround{0.0011}{4}}$\\
\morehorsp
$100\theta_\mathrm{MC}$ \vertsp ${\siround{1.04073}{5}}^{+\siround{0.00079}{5}}_{-\siround{0.00078}{5}}$\vertsp${\siround{1.04068}{5}}^{+\siround{0.00034}{5}}_{-\siround{0.00033}{5}}$\vertsp${\siround{1.04074}{5}}\pm{\siround{0.00015}{5}}$\vertsp${\siround{1.04075}{5}}\pm{\siround{0.00012}{5}}$\\
\morehorsp
$\tau$ \vertsp ${\siround{0.055}{3}}\pm{\siround{0.002}{3}}$\vertsp${\siround{0.054}{3}}\pm{\siround{0.01}{3}}$\vertsp${\siround{0.053}{3}}\pm{\siround{0.011}{3}}$\vertsp${\siround{0.055}{3}}^{+\siround{0.002}{3}}_{-\siround{0.003}{3}}$\\
\morehorsp
$H_0$ \vertsp ${\siround{67.03}{2}}\pm{\siround{2.68}{2}}$\vertsp${\siround{66.95}{2}}^{+\siround{2.65}{2}}_{-\siround{2.69}{2}}$\vertsp${\siround{66.86}{2}}^{+\siround{2.8}{2}}_{-\siround{2.69}{2}}$\vertsp${\siround{66.72}{2}}^{+\siround{2.52}{2}}_{-\siround{2.55}{2}}$\\
\morehorsp
$\Omega_K$ \vertsp ${\siround{-0.006}{3}}^{+\siround{0.007}{3}}_{-\siround{0.005}{3}}$\vertsp${\siround{-0.007}{3}}\pm{\siround{0.008}{3}}$\vertsp${\siround{-0.004}{3}}^{+\siround{0.005}{3}}_{-\siround{0.004}{3}}$\vertsp${\siround{-0.002}{3}}\pm{\siround{0.003}{3}}$\\
\morehorsp
$\log(10^{10} A_\mathrm{s})$ \vertsp ${\siround{3.093}{3}}\pm{\siround{0.01}{3}}$\vertsp${\siround{3.09}{3}}\pm{\siround{0.022}{3}}$\vertsp${\siround{3.09}{3}}\pm{\siround{0.021}{3}}$\vertsp${\siround{3.094}{3}}^{+\siround{0.005}{3}}_{-\siround{0.006}{3}}$\\
\morehorsp
$n_\mathrm{s}$ \vertsp ${\siround{0.9631}{4}}^{+\siround{0.0073}{4}}_{-\siround{0.0074}{4}}$\vertsp${\siround{0.9662}{4}}^{+\siround{0.0109}{4}}_{-\siround{0.0111}{4}}$\vertsp${\siround{0.9651}{4}}^{+\siround{0.0046}{4}}_{-\siround{0.0049}{4}}$\vertsp${\siround{0.9652}{4}}^{+\siround{0.004}{4}}_{-\siround{0.0039}{4}}$\\
\morehorsp
$w_0$ \vertsp ${\siround{-1.188}{3}}^{+\siround{0.274}{3}}_{-\siround{0.13}{3}}$\vertsp${\siround{-1.191}{3}}^{+\siround{0.254}{3}}_{-\siround{0.151}{3}}$\vertsp${\siround{-1.087}{3}}^{+\siround{0.148}{3}}_{-\siround{0.112}{3}}$\vertsp${\siround{-1.022}{3}}^{+\siround{0.088}{3}}_{-\siround{0.079}{3}}$\\
\morehorsp
$N_{\rm eff}$ \vertsp ${\siround{3.073}{3}}^{+\siround{0.244}{3}}_{-\siround{0.247}{3}}$\vertsp${\siround{3.078}{3}}^{+\siround{0.142}{3}}_{-\siround{0.14}{3}}$\vertsp${\siround{3.056}{3}}^{+\siround{0.068}{3}}_{-\siround{0.069}{3}}$\vertsp${\siround{3.058}{3}}^{+\siround{0.047}{3}}_{-\siround{0.048}{3}}$\\
\morehorsp
$m_\nu$\vertsp $< \siround{0.58}{3}$ eV\vertsp$< \siround{0.531}{3}$ eV \vertsp$< \siround{0.338}{3}$ eV\vertsp$< \siround{0.208}{3}$ eV\\\bottomrule
\end{tabular}
\caption{Forecasted constraints at $68 \%$ C.L. (upper limits at $95 \%$ C.L.) from CMB+PGWSS67 data for the experimental configurations in Table~\ref{tab:spec} in case of the $\Lambda$CDM+$\Omega_k$+$\Sigma m_{\nu}$+$N_{\rm eff}$+$w$ extended model.}
\label{CMBPGW67constraints}
\end{center}
\end{table*}

\subsection{$\Lambda$CDM+$\Omega_k$+$\Sigma m_{\nu}$+$w_a$+$w_0$ Model}

\begin{table*}[!hbtp]
\begin{center}
\begin{tabular}{lcccc}
\toprule
\horsp
Parameter \vertsp LiteBIRD+DESI \vertsp S3wide+DESI \vertsp S3deep+DESI \vertsp CMB-S4+DESI \\
\hline\hline
\morehorsp
$\Omega_\mathrm{b}h^2$ \vertsp ${\siround{0.02214}{5}}\pm{\siround{0.00018}{5}}$\vertsp${\siround{0.02218}{5}}\pm{\siround{6e-05}{5}}$\vertsp${\siround{0.02217}{5}}\pm{\siround{0.00011}{5}}$\vertsp${\siround{0.02218}{5}}\pm{\siround{3e-05}{5}}$\\
\morehorsp
$\Omega_\mathrm{c}h^2$ \vertsp ${\siround{0.1201}{4}}\pm{\siround{0.0011}{4}}$\vertsp${\siround{0.1199}{4}}\pm{\siround{0.0009}{4}}$\vertsp${\siround{0.1207}{4}}^{+\siround{0.0018}{4}}_{-\siround{0.002}{4}}$\vertsp${\siround{0.1198}{4}}\pm{\siround{0.0008}{4}}$\\
\morehorsp
$100\theta_\mathrm{MC}$ \vertsp ${\siround{1.04072}{5}}\pm{\siround{0.00049}{5}}$\vertsp${\siround{1.04075}{5}}\pm{\siround{0.00013}{5}}$\vertsp${\siround{1.0407}{5}}\pm{\siround{0.00028}{5}}$\vertsp${\siround{1.04077}{5}}\pm{\siround{0.0001}{5}}$\\
\morehorsp
$\tau$ \vertsp ${\siround{0.055}{3}}\pm{\siround{0.002}{3}}$\vertsp${\siround{0.057}{3}}\pm{\siround{0.009}{3}}$\vertsp${\siround{0.057}{3}}\pm{\siround{0.009}{3}}$\vertsp${\siround{0.055}{3}}^{+\siround{0.002}{3}}_{-\siround{0.003}{3}}$\\
\morehorsp
$H_0$ \vertsp ${\siround{66.15}{1}}\pm{\siround{2.25}{1}}$\vertsp${\siround{66.33}{1}}\pm{\siround{2.28}{1}}$\vertsp${\siround{66.35}{1}}\pm{\siround{2.42}{1}}$\vertsp${\siround{66.41}{1}}^{+\siround{2.18}{1}}_{-\siround{1.9}{1}}$\\
\morehorsp
$\Omega_K$ \vertsp ${\siround{0.0}{3}}\pm{\siround{0.002}{3}}$\vertsp${\siround{-0.0}{3}}\pm{\siround{0.002}{3}}$\vertsp${\siround{0.001}{3}}\pm{\siround{0.003}{3}}$\vertsp${\siround{-0.0}{3}}\pm{\siround{0.001}{3}}$\\
\morehorsp
$\log(10^{10} A_\mathrm{s})$ \vertsp ${\siround{3.095}{3}}\pm{\siround{0.004}{3}}$\vertsp${\siround{3.098}{3}}\pm{\siround{0.017}{3}}$\vertsp${\siround{3.1}{3}}\pm{\siround{0.018}{3}}$\vertsp${\siround{3.094}{3}}\pm{\siround{0.005}{3}}$\\
\morehorsp
$n_\mathrm{s}$ \vertsp ${\siround{0.9638}{4}}\pm{\siround{0.0042}{4}}$\vertsp${\siround{0.9644}{4}}\pm{\siround{0.0026}{4}}$\vertsp${\siround{0.9626}{4}}^{+\siround{0.006}{4}}_{-\siround{0.0059}{4}}$\vertsp${\siround{0.9645}{4}}\pm{\siround{0.0023}{4}}$\\
\morehorsp
$w_0$ \vertsp ${\siround{-0.859}{3}}^{+\siround{0.202}{3}}_{-\siround{0.259}{3}}$\vertsp${\siround{-0.883}{3}}^{+\siround{0.203}{3}}_{-\siround{0.252}{3}}$\vertsp${\siround{-0.872}{3}}^{+\siround{0.225}{3}}_{-\siround{0.269}{3}}$\vertsp${\siround{-0.901}{3}}^{+\siround{0.149}{3}}_{-\siround{0.228}{3}}$\\
\morehorsp
$w_a$ \vertsp ${\siround{-0.47}{3}}^{+\siround{0.795}{3}}_{-\siround{0.54}{3}}$\vertsp${\siround{-0.39}{3}}^{+\siround{0.749}{3}}_{-\siround{0.549}{3}}$\vertsp${\siround{-0.456}{3}}^{+\siround{0.818}{3}}_{-\siround{0.616}{3}}$\vertsp${\siround{-0.306}{3}}^{+\siround{0.661}{3}}_{-\siround{0.372}{3}}$\\
\morehorsp
$\Sigma m_\nu$\vertsp $< \siround{0.212}{3}$ eV\vertsp$< \siround{0.216}{3}$ eV\vertsp$< \siround{0.289}{3}$ eV\vertsp$< \siround{0.15}{3}$ eV\\\bottomrule
\end{tabular}
\caption{Forecasted constraints at $68 \%$ C.L. (upper limits at $95 \%$ C.L.) from CMB+DESI data for the experimental configurations in Table~\ref{tab:spec} in case of the $\Lambda$CDM+$\Omega_k$+$\Sigma m_{\nu}$+$w$+$w_a$ extended model. Note the significant increase in the error on $H_0$ (about a factor two) with respect to the $\Lambda$CDM+$\Omega_k$+$\Sigma m_{\nu}$+$N_{\rm eff}$+$w$ scenario reported before.}
\label{CMBDESIconstraints2}
\end{center}
\end{table*}

\begin{figure*}[!hbtp]
\includegraphics[width=.67\textwidth,keepaspectratio]{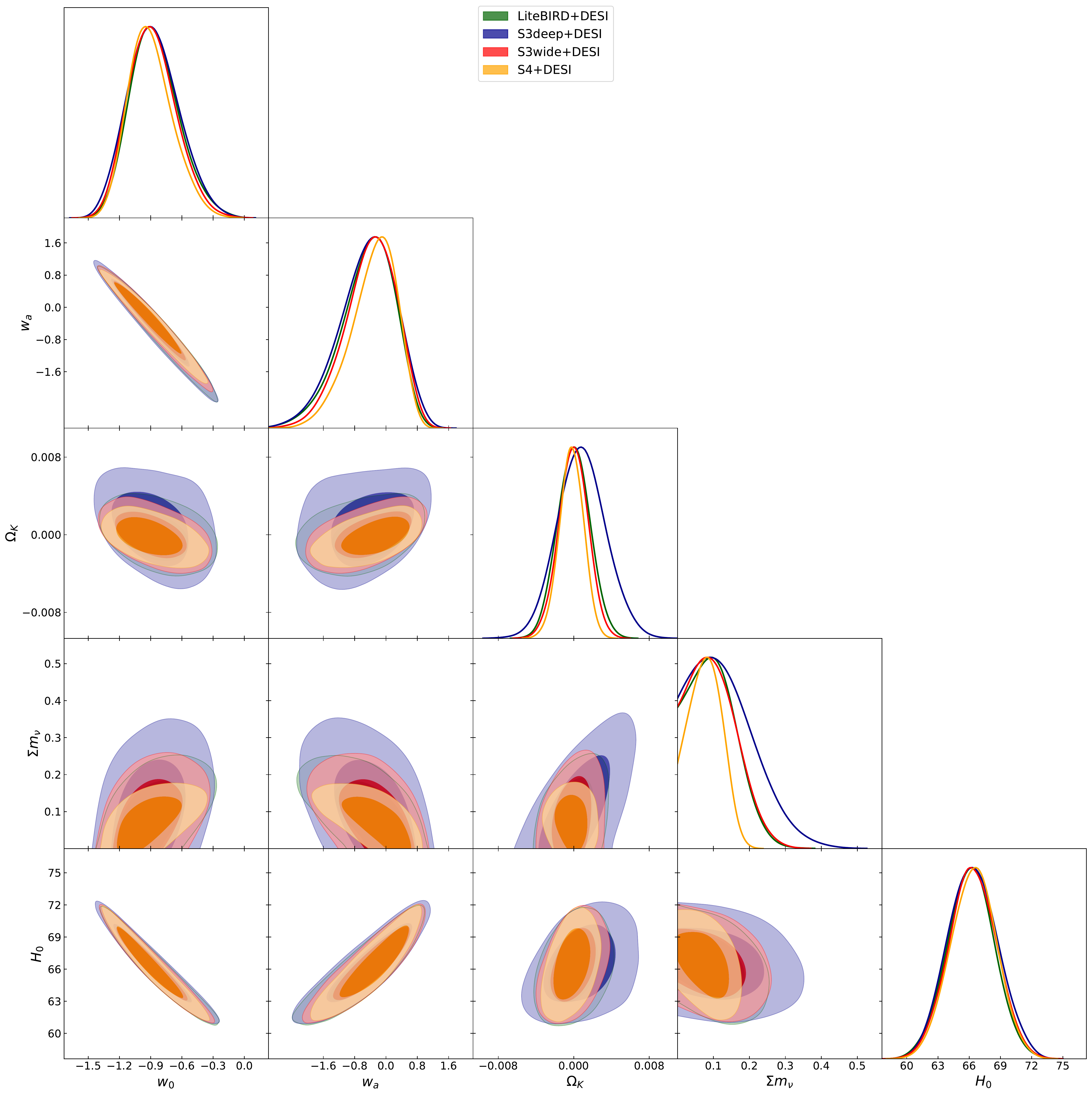}
\caption{Forecasted constraints at $68 \%$ and $95 \%$ C.L.  from CMB+DESI data for the experimental configurations in Table~\ref{tab:spec} in case of the $\Lambda$CDM+$\Omega_k$+$\Sigma m_{\nu}$+$w_0$+$w_a$ extended model.}
\label{cmbdesi2}
\end{figure*}

As shown in the previous section, the neutrino effective number $N_{\rm eff}$  will be measured with good accuracy even in extended parameter spaces. The main reason for this is due to the lack of the so-called early integrated Sachs Wolfe effect in polarization data. The inclusion of polarization helps in determining the amplitude of the EISW and $N_{\rm eff}$.

Since we are interested in evaluating the impact of a future GWSS measurement of $H_0$, it makes sense to further extend the number of geometric parameters. In what follows we substitute $N_{\rm eff}$ with $w_a$, considering therefore a dynamical dark energy equation of state described by a CPL form.

In Table~\ref{CMBDESIconstraints2} we report the constraints at $68 \%$ C.L. on cosmological parameters from the combination of future CMB and DESI data while in Figure~\ref{cmbdesi2} we report the corresponding 2D contours for the $68 \%$ and $95 \%$ confidence levels. 
If we compare with the results in Table~\ref{CMBDESIconstraints2} and in Figure~\ref{cmbdesi2} with those previously obtained assuming $w=constant$ in Table~\ref{CMBDESIconstraints} and in Figure~\ref{cmbdesi} there is now a substantial increase (about a factor two!) in the error on $H_0$. Indeed, now the combination of CMB-S4+DESI data is able to constrain the Hubble constant to only $\sim 2\,$km/s/Mpc error, i.e. to a $\sim 3 \%$ accuracy. LiteBIRD+DESI constrains $H_0$ to $\sim 3.5 \%$ accuracy.  These weaker constraints are due to the geometrical degeneracy between $H_0$, $w_a$, and $w_0$. The two dark energy parameters are now weakly determined, with uncertainties of the order of $\sim 20 \%$ for $w_0$ and $\sim 60$--70\% for $w_a$. $H_0$, $w_a$, and $w_0$ are also determined to similar accuracy by different CMB experiments, indicating that the constraining power in this case is coming primarily from DESI.
The constraint on $\Omega_k$ is virtually unchanged with respect to Table~\ref{CMBDESIconstraints}, and varies with the CMB experiment considered. The inclusion of $w_a$ weakens the future constraint on the sum of neutrino masses, $\Sigma m_{\nu}$. Other parameters, such as $n_S$, that are degenerate with $N_{\rm eff}$, are, on the contrary, now better constrained.

Given the strong degeneracy in the $w_0$--$w_a$ plane for these future experiments, it is clearly interesting to study the impact of a future GWSS determination of $H_0$. As discussed in the previous section, a $3 \%$ accuracy on $H_0$ can be reached by the HLV network after two years of operation if the BNS detection rate is $R>3500\,$Gpc$^{-3}$yr$^{-1}$, a value well inside current limits. The same accuracy can be achieved by the HLVJI network after just one year of observation even assuming the lowest BNS rate of $R=320\,$Gpc$^{-3}$yr$^{-1}$. We found that including a  $3  \%$ GWSS prior to the CMB+DESI constraints reported in Table~\ref{CMBDESIconstraints2} the constraints on $H_0$ and on the dark energy parameters could be already improved at the level of $10-30\%$.


\begin{table*}[!hbtp]
\begin{center}
\begin{tabular}{lcccc}
\toprule
\horsp
Parameter \vertsp LiteBIRD+DESI+GWSS67 \vertsp S3wide+DESI+GWSS67 \vertsp S3deep+DESI+GWSS67 \vertsp CMB-S4+DESI+GWSS67 \\
\hline\hline
\morehorsp
$\Omega_\mathrm{b}h^2$ \vertsp ${\siround{0.02214}{5}}\pm{\siround{0.00017}{5}}$\vertsp${\siround{0.02218}{5}}\pm{\siround{5e-05}{5}}$\vertsp${\siround{0.02217}{5}}\pm{\siround{0.00012}{5}}$\vertsp${\siround{0.02218}{5}}\pm{\siround{3e-05}{5}}$\\
\morehorsp
$\Omega_\mathrm{c}h^2$ \vertsp ${\siround{0.1202}{4}}^{+\siround{0.001}{4}}_{-\siround{0.0011}{4}}$\vertsp${\siround{0.12}{4}}\pm{\siround{0.0009}{4}}$\vertsp${\siround{0.1207}{4}}^{+\siround{0.0017}{4}}_{-\siround{0.002}{4}}$\vertsp${\siround{0.1198}{4}}\pm{\siround{0.0008}{4}}$\\
\morehorsp
$100\theta_\mathrm{MC}$ \vertsp ${\siround{1.04074}{5}}\pm{\siround{0.00048}{5}}$\vertsp${\siround{1.04075}{5}}\pm{\siround{0.00013}{5}}$\vertsp${\siround{1.0407}{5}}\pm{\siround{0.00028}{5}}$\vertsp${\siround{1.04077}{5}}\pm{\siround{0.0001}{5}}$\\
\morehorsp
$\tau$ \vertsp ${\siround{0.055}{3}}\pm{\siround{0.002}{3}}$\vertsp${\siround{0.057}{3}}\pm{\siround{0.008}{3}}$\vertsp${\siround{0.057}{3}}\pm{\siround{0.009}{3}}$\vertsp${\siround{0.055}{3}}\pm{\siround{0.002}{3}}$\\
\morehorsp
$H_0$ \vertsp ${\siround{67.21}{2}}^{+\siround{0.62}{2}}_{-\siround{0.63}{2}}$\vertsp${\siround{67.23}{2}}^{+\siround{0.67}{2}}_{-\siround{0.63}{2}}$\vertsp${\siround{67.24}{2}}\pm{\siround{0.64}{2}}$\vertsp${\siround{67.23}{2}}^{+\siround{0.63}{2}}_{-\siround{0.64}{2}}$\\
\morehorsp
$\Omega_K$ \vertsp ${\siround{0.0}{3}}\pm{\siround{0.002}{3}}$\vertsp${\siround{0.0}{3}}\pm{\siround{0.001}{3}}$\vertsp${\siround{0.001}{3}}\pm{\siround{0.002}{3}}$\vertsp${\siround{-0.0}{3}}\pm{\siround{0.001}{3}}$\\
\morehorsp
$\log(10^{10} A_\mathrm{s})$ \vertsp ${\siround{3.095}{3}}\pm{\siround{0.004}{3}}$\vertsp${\siround{3.098}{3}}^{+\siround{0.016}{3}}_{-\siround{0.017}{3}}$\vertsp${\siround{3.1}{3}}\pm{\siround{0.018}{3}}$\vertsp${\siround{3.095}{3}}\pm{\siround{0.005}{3}}$\\
\morehorsp
$n_\mathrm{s}$ \vertsp ${\siround{0.9638}{4}}\pm{\siround{0.0043}{4}}$\vertsp${\siround{0.9642}{4}}\pm{\siround{0.0026}{4}}$\vertsp${\siround{0.9625}{4}}\pm{\siround{0.0058}{4}}$\vertsp${\siround{0.9644}{4}}\pm{\siround{0.0022}{4}}$\\
\morehorsp
$w_0$ \vertsp ${\siround{-0.974}{3}}^{+\siround{0.078}{3}}_{-\siround{0.089}{3}}$\vertsp${\siround{-0.978}{3}}^{+\siround{0.081}{3}}_{-\siround{0.089}{3}}$\vertsp${\siround{-0.969}{3}}^{+\siround{0.084}{3}}_{-\siround{0.095}{3}}$\vertsp${\siround{-0.985}{3}}^{+\siround{0.066}{3}}_{-\siround{0.082}{3}}$\\
\morehorsp
$w_a$ \vertsp ${\siround{-0.147}{3}}^{+\siround{0.377}{3}}_{-\siround{0.282}{3}}$\vertsp${\siround{-0.127}{3}}^{+\siround{0.36}{3}}_{-\siround{0.304}{3}}$\vertsp${\siround{-0.188}{3}}^{+\siround{0.404}{3}}_{-\siround{0.32}{3}}$\vertsp${\siround{-0.08}{3}}^{+\siround{0.319}{3}}_{-\siround{0.225}{3}}$\\
\morehorsp
$\Sigma m_\nu$\vertsp $< \siround{0.196}{3}$ eV\vertsp$< \siround{0.205}{3}$ eV\vertsp$< \siround{0.278}{3}$ eV\vertsp$< \siround{0.14}{3}$ eV\\\bottomrule
\end{tabular}
\caption{Forecasted constraints at $68 \%$ C.L. (upper limits at $95 \%$ C.L.)  from CMB+DESI+GWSS67 data for the experimental configurations in Table~\ref{tab:spec} in case of the $\Lambda$CDM+$\Omega_k$+$\Sigma m_{\nu}$+$w$+$w_a$ extended model. Note the significant improvement in accuracy on $H_0$ and on the dark energy parameters $w_0$ and $w_a$ with respect to the CMB+DESI case.}
\label{CMBDESIGW672}
\end{center}
\end{table*}

\begin{figure*}[!hbtp]
\includegraphics[width=.67\textwidth,keepaspectratio]{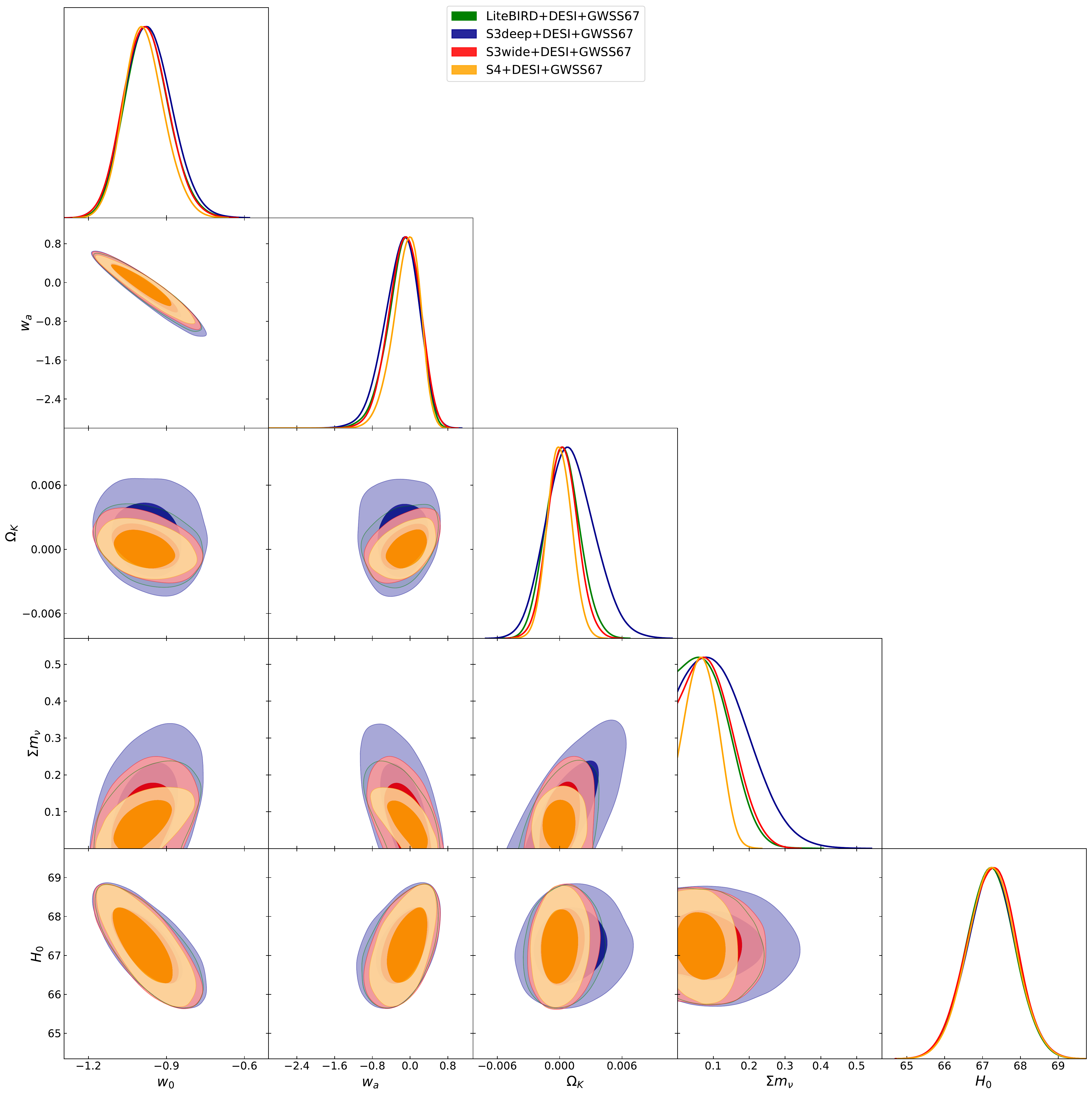}
\caption{Forecasted constraints at $68 \%$ and $95 \%$ C.L. from CMB+DESI+GWSS67 data for the experimental configurations in Table~\ref{tab:spec} in case of the $\Lambda$CDM+$\Omega_k$+$\Sigma m_{\nu}$+$w_0$+$w_a$ extended model.}
\label{cmbdesigw2}
\end{figure*}

However, a $\sim 1\%$ accuracy on $H_0$ is also directly attainable by future GWSS measurements, and it is interesting to discuss the impact of this improved determination on future combined cosmological parameter measurements.
We  report the constraints on cosmological parameters  for CMB+DESI+GWSS67 in Table~\ref{CMBDESIGW672} and the corresponding 2D confidence levels in Figure~\ref{cmbdesigw2}.
The measured value of the Hubble constant is practically identical to the assumed prior from the standard sirens (GWSS67), indicating that the standard siren measurements are contributing to the combined constraints on all related cosmological parameters.
In particular, the constraints on the dark energy parameters $w_0$ and $w_a$ are substantially improved, by a factor $\sim 1.6$--2.8, with the inclusion of the standard siren measurements.

Finally, in Table~\ref{CMBDESIPGW672} we report the expected constraints when combining future CMB data with a, pessimistic, PGWSS67 prior on the Hubble parameter. As we can see, including the PGWSS67 prior will improve the constraints on the dark energy parameters by $\sim 20-30 \%$ respect to CMB+DESI data. A $\sim 4 \%$ determination of the Hubble parameter can be therefore useful in this theoretical framework even when considering the CMB+DESI dataset.
However the constraints achievable with the PGWSS67 prior on $w_0$ will be about a factor two larger than those achievable with the GWSS67 prior.

\begin{table*}[!hbtp]
\begin{center}
\begin{tabular}{lcccc}
\toprule
\horsp
Parameter \vertsp LiteBIRD+DESI+PGWSS67\vertsp S3wide+DESI+PGWSS67\vertsp S3deep+DESI+PGWSS67\vertsp CMB-S4+DESI+PGWSS67\\
\hline\hline
\morehorsp
$\Omega_\mathrm{b}h^2$ \vertsp ${\siround{0.02214}{5}}\pm{\siround{0.00017}{5}}$\vertsp${\siround{0.02217}{5}}\pm{\siround{0.00011}{5}}$\vertsp${\siround{0.02218}{5}}\pm{\siround{6e-05}{5}}$\vertsp${\siround{0.02218}{5}}\pm{\siround{3e-05}{5}}$\\
\morehorsp
$\Omega_\mathrm{c}h^2$ \vertsp ${\siround{0.1201}{4}}^{+\siround{0.001}{4}}_{-\siround{0.0011}{4}}$\vertsp${\siround{0.1207}{4}}^{+\siround{0.0018}{4}}_{-\siround{0.002}{4}}$\vertsp${\siround{0.1199}{4}}\pm{\siround{0.0009}{4}}$\vertsp${\siround{0.1199}{4}}\pm{\siround{0.0008}{4}}$\\
\morehorsp
$100\theta_\mathrm{MC}$ \vertsp ${\siround{1.04073}{5}}\pm{\siround{0.00048}{5}}$\vertsp${\siround{1.0407}{5}}\pm{\siround{0.00028}{5}}$\vertsp${\siround{1.04075}{5}}\pm{\siround{0.00013}{5}}$\vertsp${\siround{1.04076}{5}}\pm{\siround{0.0001}{5}}$\\
\morehorsp
$\tau$ \vertsp ${\siround{0.055}{3}}\pm{\siround{0.002}{3}}$\vertsp${\siround{0.057}{3}}\pm{\siround{0.009}{3}}$\vertsp${\siround{0.057}{3}}\pm{\siround{0.008}{3}}$\vertsp${\siround{0.055}{3}}^{+\siround{0.002}{3}}_{-\siround{0.003}{3}}$\\
\morehorsp
$H_0$ \vertsp ${\siround{66.64}{2}}^{+\siround{1.71}{2}}_{-\siround{1.72}{2}}$\vertsp${\siround{66.76}{2}}^{+\siround{1.82}{2}}_{-\siround{1.83}{2}}$\vertsp${\siround{66.74}{2}}^{+\siround{1.74}{2}}_{-\siround{1.72}{2}}$\vertsp${\siround{66.79}{2}}^{+\siround{1.69}{2}}_{-\siround{1.71}{2}}$\\
\morehorsp
$\Omega_K$ \vertsp ${\siround{0.0}{3}}\pm{\siround{0.002}{3}}$\vertsp${\siround{0.001}{3}}^{+\siround{0.002}{3}}_{-\siround{0.003}{3}}$\vertsp${\siround{0.0}{3}}\pm{\siround{0.001}{3}}$\vertsp${\siround{-0.0}{3}}\pm{\siround{0.001}{3}}$\\
\morehorsp
$\log(10^{10} A_\mathrm{s})$ \vertsp ${\siround{3.095}{3}}\pm{\siround{0.004}{3}}$\vertsp${\siround{3.1}{3}}\pm{\siround{0.018}{3}}$\vertsp${\siround{3.098}{3}}\pm{\siround{0.017}{3}}$\vertsp${\siround{3.095}{3}}\pm{\siround{0.005}{3}}$\\
\morehorsp
$n_\mathrm{s}$ \vertsp ${\siround{0.9638}{4}}\pm{\siround{0.0042}{4}}$\vertsp${\siround{0.9625}{4}}^{+\siround{0.0059}{4}}_{-\siround{0.0058}{4}}$\vertsp${\siround{0.9643}{4}}\pm{\siround{0.0026}{4}}$\vertsp${\siround{0.9643}{4}}\pm{\siround{0.0022}{4}}$\\
\morehorsp
$w_0$ \vertsp ${\siround{-0.912}{3}}^{+\siround{0.159}{3}}_{-\siround{0.195}{3}}$\vertsp${\siround{-0.918}{3}}^{+\siround{0.172}{3}}_{-\siround{0.204}{3}}$\vertsp${\siround{-0.926}{3}}^{+\siround{0.163}{3}}_{-\siround{0.19}{3}}$\vertsp${\siround{-0.938}{3}}^{+\siround{0.142}{3}}_{-\siround{0.182}{3}}$\\
\morehorsp
$w_a$ \vertsp ${\siround{-0.323}{3}}^{+\siround{0.619}{3}}_{-\siround{0.444}{3}}$\vertsp${\siround{-0.33}{3}}^{+\siround{0.643}{3}}_{-\siround{0.497}{3}}$\vertsp${\siround{-0.271}{3}}^{+\siround{0.588}{3}}_{-\siround{0.46}{3}}$\vertsp${\siround{-0.213}{3}}^{+\siround{0.548}{3}}_{-\siround{0.375}{3}}$\\
\morehorsp
$m_\nu$\vertsp $< \siround{0.205}{3}$ eV\vertsp$< \siround{0.284}{3}$ eV\vertsp$< \siround{0.211}{3}$ eV\vertsp$< \siround{0.148}{3}$ eV\\\bottomrule
\end{tabular}
\caption{Forecasted constraints at $68 \%$ C.L. (upper limits at $95 \%$ C.L.)  from CMB+DESI+GWSS67 data for the experimental configurations in Table~\ref{tab:spec} in case of the $\Lambda$CDM+$\Omega_k$+$\Sigma m_{\nu}$+$w$+$w_a$ extended model. Note the significant improvement in accuracy on $H_0$ and on the dark energy parameters $w_0$ and $w_a$ with respect to the CMB+DESI case.}
\label{CMBDESIPGW672}
\end{center}
\end{table*}

\subsection{Figure of Merit} 

\begin{table*}[!hbtp]
\begin{center}
\begin{tabular}{|c|c|cccc|}
\hline
\morehorsp
Model \vertsp    Dataset     \vertsp\  LiteBIRD \vertsp S3deep \vertsp S3wide \vertsp CMB-S4 \\  
\hline
\hline
\morehorsp
$\Lambda$CDM+$\Omega_k$+$\Sigma m_{\nu}$+$N_{\rm eff}$+$w$  & CMB & $5$ & $1$& $398$ &$29236$ \\
\horsp
\vertsp CMB$+$PGWSS67 & $110$ & $40$& $12732$ &$2.2\times10^{6}$ \\
\horsp
\vertsp CMB$+$GWSS67 & $262$ & $104$& $50929$ &$1.2\times10^{7}$ \\
\horsp
\vertsp CMB$+$DESI & $6659$ & $2415$& $383240$ & $3.74\times10^{7}$ \\
\horsp
\vertsp CMB$+$DESI$+$PGWSS67 & $7735$ & $2807$& $422008$ & $4.06\times10^{7}$\\
\horsp
\vertsp CMB$+$DESI$+$GWSS67 & $16928$ & $5484$& $752879$ & $7.39\times10^{7}$\\
\hline
\hline
\morehorsp
$\Lambda$CDM+$\Omega_k$+$\Sigma m_{\nu}$+$w_0$+$w_a$ & CMB & $7$ & $1$& $170$ &$9223$ \\
\horsp
\vertsp CMB$+$PGWSS67 & $111$ & $18$& $2732$ &$14402$ \\
\horsp
\vertsp CMB$+$GWSS67 & $291$ & $43$& $9231$ &$589791$ \\
\horsp
\vertsp CMB$+$DESI \vertsp$13335$ & $2394$& $227590$ &$1.04\times10^{7}$\\
\horsp
\vertsp CMB$+$DESI$+$PGWSS67 & $19458$ & $3577$& $323789$ &$1.6\times10^{7}$\\
\horsp
\vertsp CMB$+$DESI$+$GWSS67 & $57928$ & $11735$& $1.01\times10^{6}$ &$5.7\times10^{7}$\\
\hline
\end{tabular}
\end{center}
\caption{Improvement with respect to simulated CMB data of the global Figure of Merit for the two theoretical scenarios considered in the paper and for different combination of datasets. The FoM is normalized to the S3deep CMB alone case that provides the less constraining results.}
\label{tab:FOMs}
\end{table*}

It is interesting to quantify the improvement of a GWSS prior by comparing the overall Figure of Merit for the cases considered.
Given an experimental configuration and a set of $N$ parameters $p_i$ with $i=(1,...N)$, we can define the FoM  from the covariance matrix of uncertainties on $p_i$ as (see e.g. 
\cite{wmap9,core1}):

\begin{equation}
\rm FoM = (\det[\mathrm{cov} \ p_i\}])^{-1/2}
\end{equation}

\noindent that is proportional to the inverse of the volume of the constrained parameters space. It is important to stress that this FoM considers the whole parameter space and not just the dark energy parameters as in \cite{weinberg}.

In Table~\ref{tab:FOMs} we report the FoM for the two theoretical scenarios considered in this paper and for different combinations of datasets. The FoM are normalized to the S3deep, CMB only, value.
As we can see, in the case of $\Lambda$CDM+$\Omega_k$+$\Sigma m_{\nu}$+$N_{\rm eff}$+$w$ there is a significant improvement in FoM when the GWSS67 prior is included with the CMB data. The improvement is significant (between a factor $\sim 50$ and $\sim 400$) and larger  in the case of the CMB-S4 dataset. A smaller but still significant improvement is present when the PGWSS67 prior is considered. This clearly shows that, once the geometrical degeneracies are broken by the introduction of the GWSS prior, there is a significantly improved parameter determination with this dataset.
It is interesting also to note that the S3wide configuration has a constraining power that is superior to LiteBIRD+GWSS67 and S3deep+GWSS67. When the DESI dataset is included there is an improvement by a factor $\sim 1000$ and $\sim 2400$. In this case the CMB dataset that would better benefit by the inclusion of the DESI data is S3deep. Both S3deep+DESI and LiteBIRD+DESI have a smaller FoM than S3wide+GWSS67, and S3wide+DESI has less constraining power than CMB-S4+GWSS67.
When further including the GWSS67 prior the improvement in FoM is about a factor 2--3 with respect to the CMB+DESI case, clearly showing that GWSS will be useful in further constraining the parameter space.
However, when considering the more pessimistic PGWSS67 prior the improvement with respect to the CMB+DESI case is just $\sim 10-20 \%$.

In the case of the $\Lambda$CDM+$\Omega_k$+$\Sigma m_{\nu}$+$w_0$+$w_a$ model the improvement in the FoM obtained by the inclusion of the GWSS67 prior in the case of the CMB data is about a factor of $\sim 50$. With the DESI dataset the improvement is a factor of $\sim 1000$--2400. As we can see these improvements are smaller if compared to the $\Lambda$CDM+$\Omega_k$+$\Sigma m_{\nu}$+$N_{\rm eff}$+$w$ scenario, showing that in this case the parameter degeneracies are more severe. When the GWSS67 prior is included the improvement is about a factor $\sim 4$--6, larger if compared with the similar data combination for the $\Lambda$CDM+$\Omega_k$+$\Sigma m_{\nu}$+$N_{\rm eff}$+$w$ scenario. The combination of LiteBIRD, S3deep, and S3wide with DESI data has less constraining power than CMB-S4+GWSS67.
The inclusion of a PGWSS67 prior can improve by a $\sim 60 \%$ the FoM of CMB-S4 an CMB-S4+DESI.

\begin{figure*}[!hbtp]
\includegraphics[width=.47\textwidth,keepaspectratio]{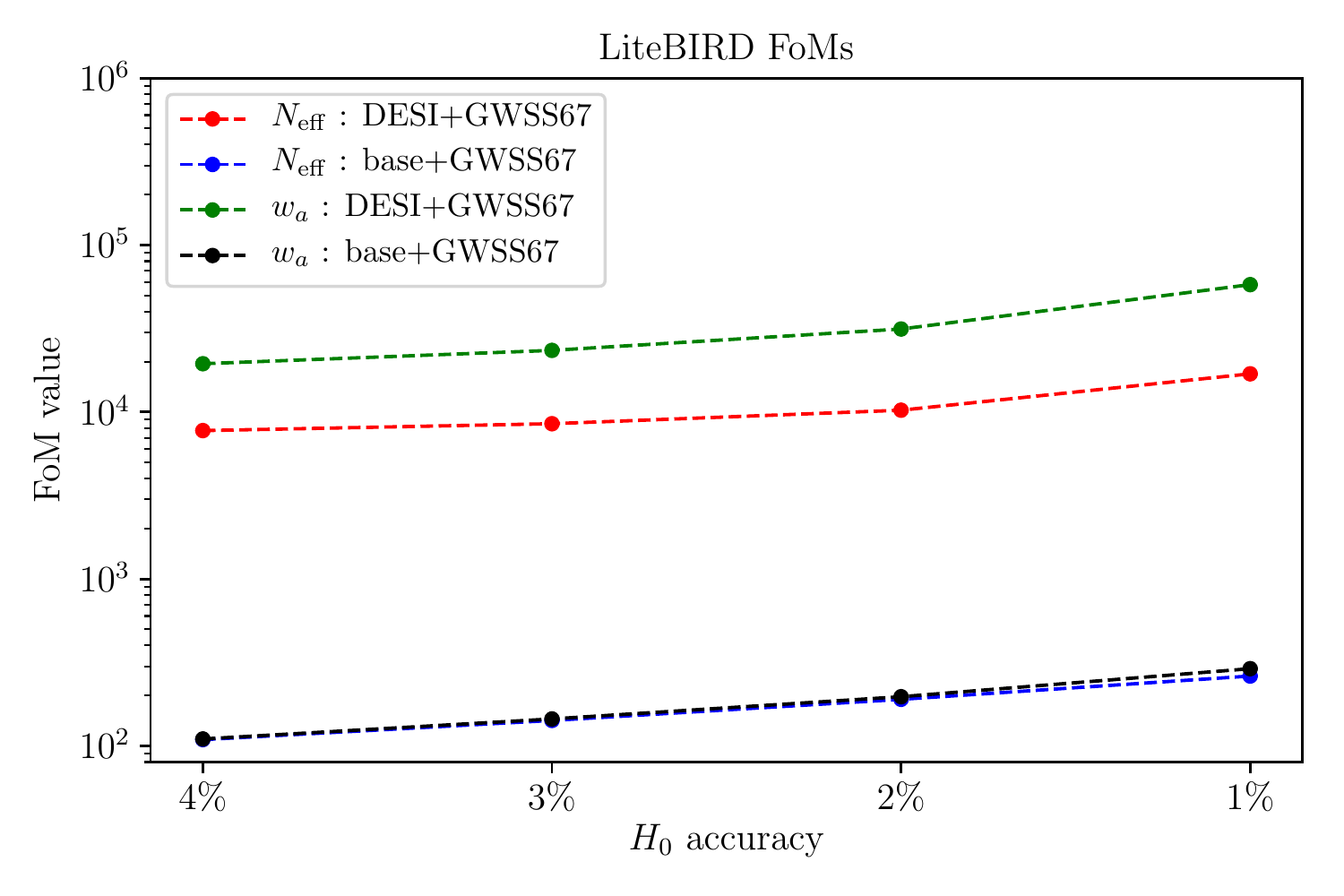}
\includegraphics[width=.47\textwidth,keepaspectratio]{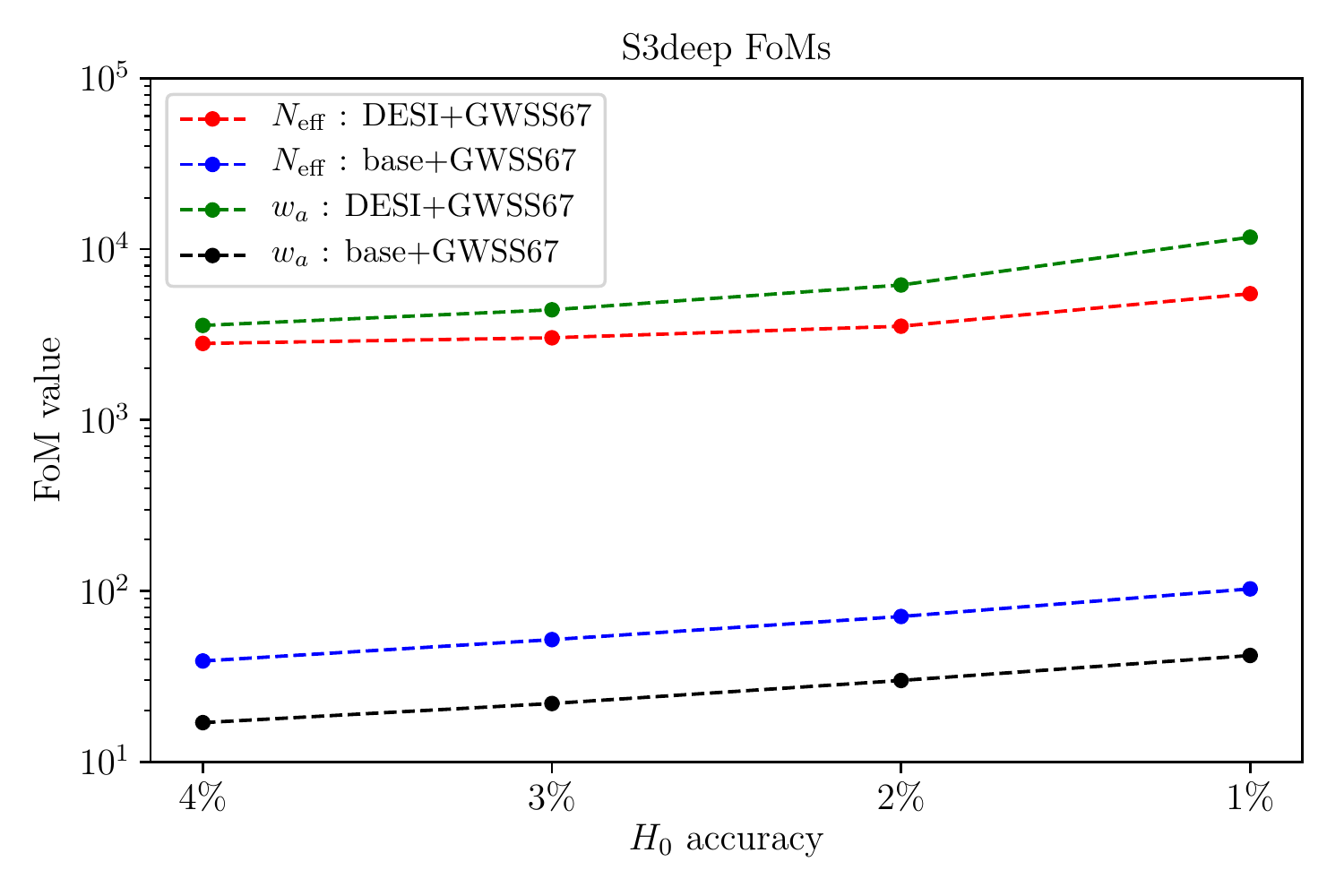}
\includegraphics[width=.47\textwidth,keepaspectratio]{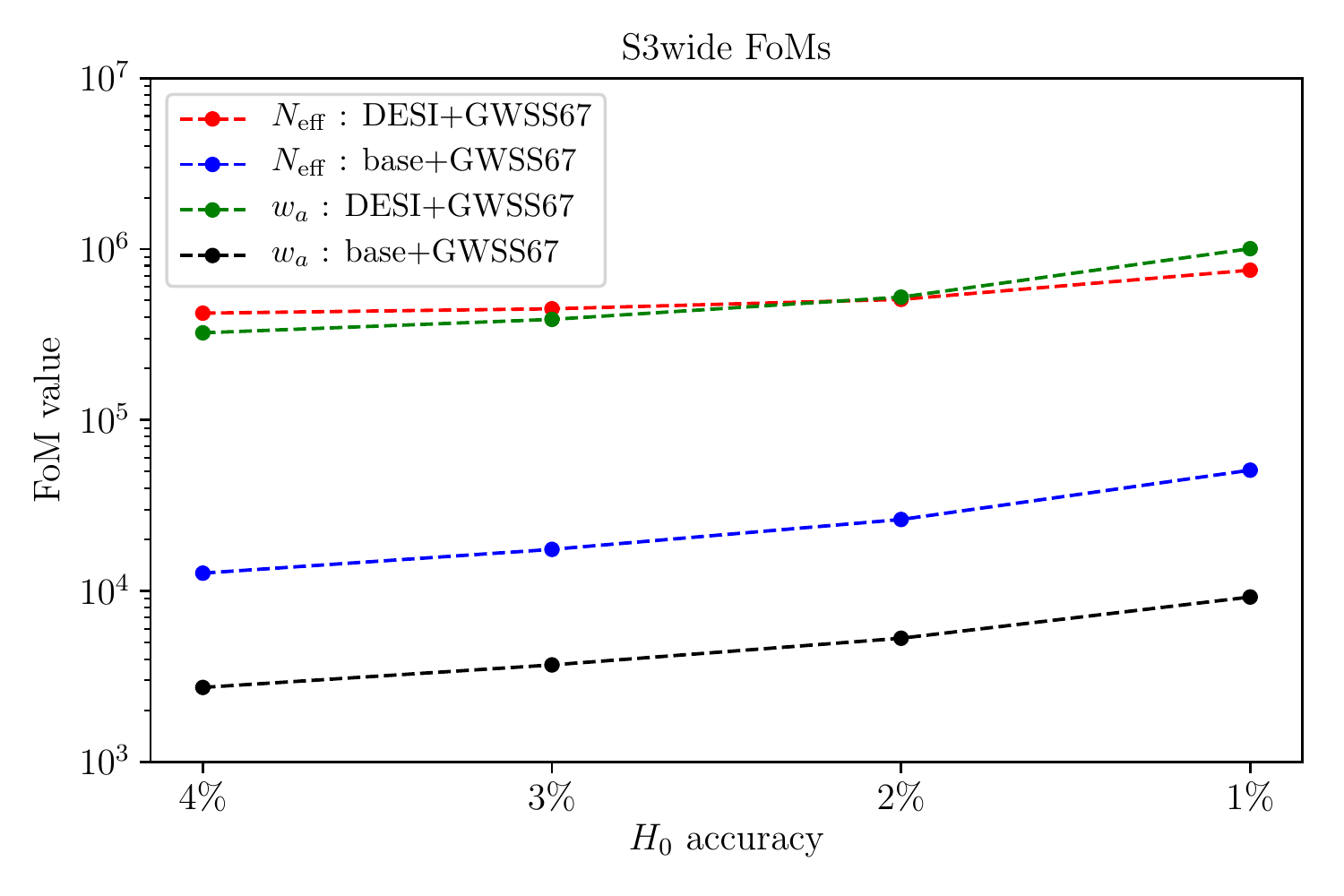}
\includegraphics[width=.47\textwidth,keepaspectratio]{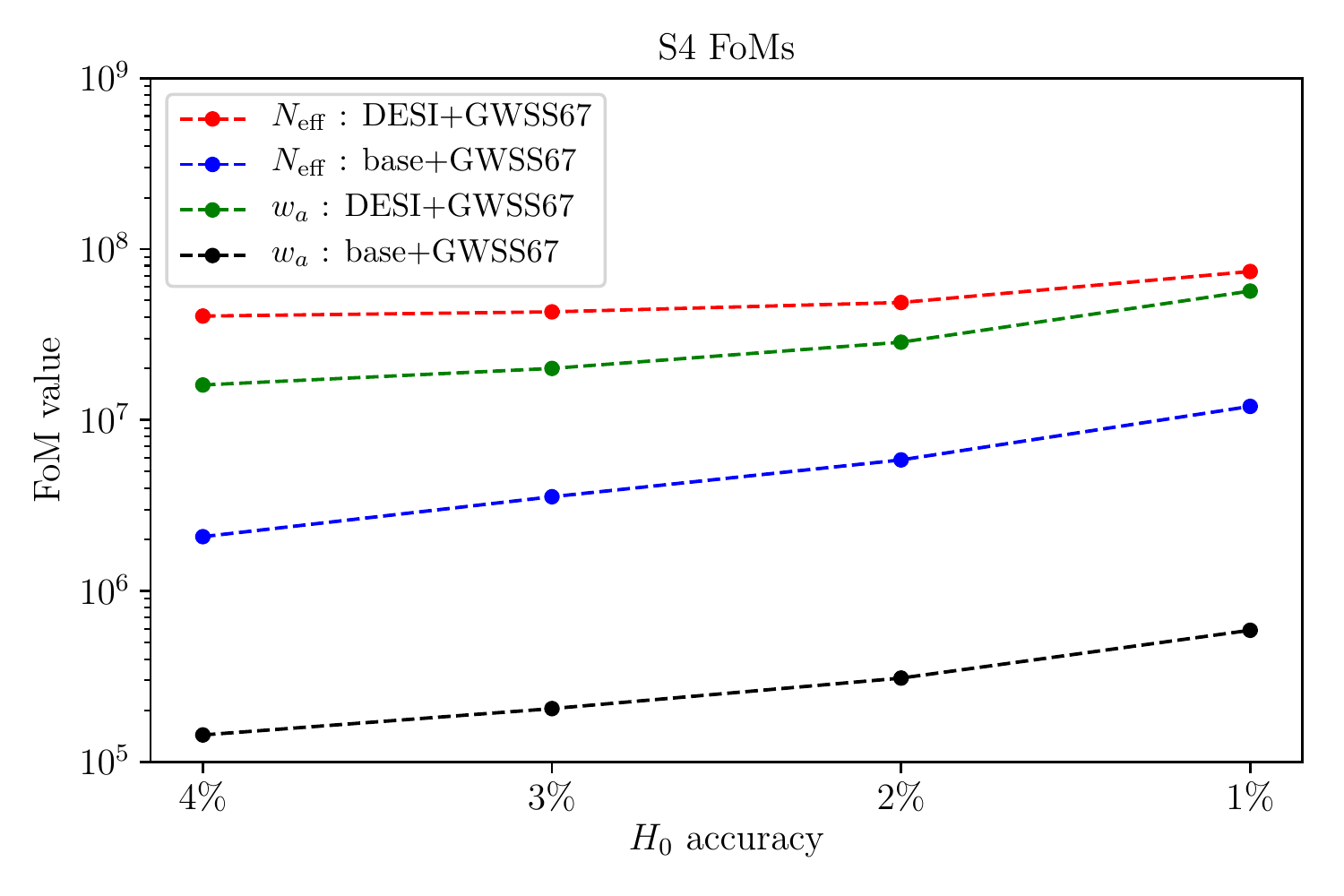}
\caption{Figures of Merit for the theoretical models and experimental configurations considered in function of different priors on the Hubble parameter with a $4\%$, $3\%$, $2\%$, and $1\%$ accuracy respectively. The assumed CMB datasets are LiteBIRD (Top Left), S3deep (Top Right), S3wide (Bottom Left), and CMB-S4 (Bottom Right).}
\label{fomall}
\end{figure*}

Finally, in order to better visualize the impact of a future prior on $H_0$,  we plot in Figure ~\ref{fomall} the values of the FoM in function of of $4$ different expected accuracies on the Hubble constant: $4 \%$, $3 \%$, $2 \%$, and $1 \$$.
We can firstly clearly see that the FoM will be in general larger in case of the "$w_0+w_a$" scenario with respect to the "$w_0+N_{\rm eff}$" for any experimental configuration (with the exception of LiteBIRD). The inclusion of an external prior on the Hubble parameter is therefore more efficient in improving the constraints in the case of a "$w_0+w_a$" model, where dynamical dark energy is considered. Secondly, while in the CMB only scenario an improvement in the accuracy of $H_0$ is always reflected in a substantial increase in the FoM, it seems that in the case of CMB+DESI and for the "$w_0+N_{\rm eff}$" model (the red lines in the figure) a significant increase is expected when moving to an accuracy below $2 \%$. An improved accuracy in $H_0$ from $4 \%$ to $2 \%$ produces larger improvements in the FoM for the CMB+DESI dataset in the case of the "$w_0+w_a$" scenario.

\section{Conclusions}

The recent observations of gravitational waves and electromagnetic emission produced by the merger of the binary neutron-star system GW170817 has introduced a complementary and direct method for measuring the Hubble constant. In the coming decade GW standard sirens are expected to produce constraints on $H_0$ with $\sim 1 \%$ accuracy. At the same time, improved constraints are expected from CMB experiments and from BAO surveys. 
In an extended $\Lambda$CDM parameter space, where we have considered variations  in curvature, neutrino mass, and the dark energy equation of state, we have found that a combination of future CMB and BAO data can constraint the Hubble constant at the level of 1.5--2\%.
A similar accuracy may be reached by the HLV network in the second year of observations if the the BNS rate is $R\ge2800$ Gpc$^{-3}$yr$^{-1}$, in agreement with current limits on $R$, or by the HLVJI network after one year of observations with a more conservative BNS detection rate of $R\ge 1540$ Gpc$^{-3}$yr$^{-1}$.

Gravitational wave standard sirens may reach a 1\% measurement of $H_0$ within the decade, which when combined with future CMB data would constrain
curvature to $0.3 \%$ and the dark energy equation of state to $\sim 5 \%$. A GWSS measurement of the Hubble constant would also improve the constraints on these geometrical parameters coming from future CMB+BAO data by by 30--40\%. In addition, the current 2$\sigma$ Hubble tension between CMB+BAO and supernova data could be strengthened to $5\sigma$ with the inclusion of standard siren constraints.

When we further include time variations in the dark energy equation of state, parameterizing its evolution with a CPL function, we find that future CMB+BAO data will constrain the Hubble constant to $\sim 3 \%$. This level of accuracy on $H_0$ can be independently reached by the HLV network of interferometers after the second year of operation if the BNS detection rate is $R>3500$ Gpc$^{-3}$yr$^{-1}$, a value again well inside current limits, or by the HLVJI network after one year of observations even considering a low BNS detection rate of $R=320$ Gpc$^{-3}$yr$^{-1}$. This standard siren measurement would therefore improve the CMB+BAO constraints on this model at the level of $10-30\%$.

Assuming a future $H_0$ accuracy of $\sim 1 \%$ from standard sirens, as to be expected within the decade, we find that the constraints on the dark energy equation of state parameters $w_0$ and $w_a$ from future CMB+BAO datasets can be improved by a factor 1.6--2.8. We conclude that standard siren measurements by the HLV and HLVJI gravitational-wave detector networks over the coming decade may significantly improve our understanding of cosmology. 

We have also found that even a more pessimistic determination of $H_0$, with a a $\sim 4\%$ accuracy can significantly improve the constraints from CMB alone data in case of a $\Lambda$CDM+$\Omega_k$+$\Sigma m_{\nu}$+$N_{\rm eff}$+$w$ model and from CMB alone and CMB+DESI data
in case of a $\Lambda$CDM+$\Omega_k$+$\Sigma m_{\nu}$+$w_0$+$w_a$
model.

Finally it is clearly worth mentioning that similar constraints on $H_0$ and dark energy parameters could come by combining CMB and BAO data with other complementary probes such as supernovae and cosmic shear (see e.g. \cite{aubourg,gwss11}). In this case future constraints from GWSS will play a crucial role in confirming these results and cross-validating the different approaches. In addition, these comparisons offer the exciting possibility of discovering new physics beyond the $\Lambda$-CDM scenario.

\acknowledgments 
 
EDV acknowledges support from the European Research Council in the form of a Consolidator Grant with number 681431. AM thanks the University of Manchester and the Jodrell Bank Center for Astrophysics for hospitality. AM and FR are supported by TASP, iniziativa specifica INFN. DEH was partially supported by NSF grant PHY-1708081. He was also supported by the Kavli Institute for Cosmological Physics at the University of Chicago through NSF grant PHY-1125897 and an endowment from the Kavli Foundation. DEH also gratefully acknowledges support from the Marion and Stuart Rice Award. We thank Cristiano Palomba for useful comments.

\end{document}